\documentclass[fleqn,usenatbib]{mnras}

\usepackage{newtxtext,newtxmath}
\usepackage[T1]{fontenc}
\usepackage{ae,aecompl}

\usepackage{graphicx}	% Including figure files
\usepackage{amsmath}	% Advanced maths commands
\usepackage{multirow}

\title[Morphometry as a probe of the evolution of jellyfish galaxies]{Morphometry as a probe of the evolution of jellyfish galaxies: evidence of broadening in the surface brightness profiles of ram-pressure stripping candidates in the multi-cluster system A901/A902}

\author[Roman-Oliveira et al.]{Fernanda Roman-Oliveira,$^{1, 2}$\thanks{E-mail: romanoliveira@astro.rug.nl}
Ana L. Chies-Santos$^{2}$,
Fabricio Ferrari$^{3}$,
\newauthor Geferson Lucatelli$^{3}$,
Bruno Rodr\'iguez Del Pino$^{4}$
\\
$^{1}$Kapteyn Astronomical Institute, University of Groningen, Landleven 12, 9747AD, Groningen, the Netherlands \\
$^{2}$Departamento de Astronomia, Instituto de F\'isica, Universidade Federal do Rio Grande do Sul, Porto Alegre, RS, Brazil\\
$^{3}$Instituto de Matem\'atica, Estat\'istica e F\'isica, Universidade Federal de Rio Grande, Rio Grande, RS, Brazil\\
$^{4}$Centro de Astrobiolog\'ia (CSIC-INTA), Torrej\'on de Ardoz, Madrid, Spain
}

\date{Accepted XXX. Received YYY; in original form ZZZ}

\pubyear{2020}

\begin{document}
\label{firstpage}
\pagerange{\pageref{firstpage}--\pageref{lastpage}}
\maketitle

\begin{abstract}%ok
We explore the morphometric properties of a group of 73 ram pressure stripping candidates in the A901/A902 multi-cluster system, at z$\sim$0.165, to characterise the morphologies and structural evolution of jellyfish galaxies.
By employing a quantitative measurement of morphometric indicators with the algorithm \textsc{morfometryka} on Hubble Space Telescope (F606W) images of the galaxies, we present a novel morphology-based method for determining trail vectors.
We study the surface brightness profiles and curvature of the candidates and compare the results obtained with two analysis packages, \textsc{morfometryka} and \textsc{iraf/ellipse} on retrieving information of the irregular structures present in the galaxies. Our morphometric analysis shows that the ram pressure stripping candidates have peculiar concave regions in their surface brightness profiles. Therefore, these profiles are less concentrated (lower S\'ersic indices) than other star forming galaxies that do not show morphological features of ram pressure stripping.
In combination with morphometric trail vectors, this feature could both help identify galaxies undergoing ram-pressure stripping and reveal spatial variations in the star formation rate.
\end{abstract}

\begin{keywords}
galaxies: clusters: intracluster medium -- galaxies: evolution --  galaxies: irregular
\end{keywords}

%%%%%%%%%%%%%%%%%%%%%%%%%%%%%%%%%%%%%%%%%%%%%%%%%%

%%%%%%%%%%%%%%%%% BODY OF PAPER %%%%%%%%%%%%%%%%%%

\section{Introduction}%ok
Previous research shows that dense environments influence the evolution of galaxies \citep{dressler80, butcheroemler84}.
Passive elliptical galaxies are more frequently found in the centre of galaxy clusters and star forming disc galaxies are more common as satellite galaxies \citep{bamford09}. This is linked to transformations in both morphology and galaxy properties, such as colours and star formation rates.
%tidal stripping, ram pressure stripping, minor/major mergers,
What is yet not clear is the impact of the several external galaxy evolution drivers concurrently at play in such environments, e.g. stripping through tidal \citep{barnes92} and ram pressure interactions \citep{gunngott72}, galaxy harassment \citep{moore96}, mergers \citep{bekki99, barnes92}, starvation or strangulation \citep{larson80}.
In this work, we explore the relationship between the ram pressure stripping effect in galaxies and their evolution in galaxy clusters.

Ram pressure stripping is an efficient mechanism in removing gas from orbiting galaxies in clusters. It occurs when there is a hydrodynamic friction between the interstellar medium (ISM) in a galaxy and the intracluster medium (ICM) as the galaxy falls into a galaxy cluster. 
Jellyfish galaxies are the most representative example of galaxies undergoing ram pressure stripping, these are rare and extreme cases of galaxies with extensive tails that can be identified throughout many wavelengths \citep{poggianti19b}.
Many studies over the past decade provide important information on the origins, distribution and physical properties of ram pressure stripped galaxies \citep{poggianti16, ebeling14, smith10}.
Recently, there have been new statistically significant studies on the properties of large samples of jellyfish galaxies such as the GaSP collaboration \citep{poggianti17}, the \cite{mcpartland16} sample in massive clusters and the rich population of ram pressure stripping candidates found in the Abell 901/2 system as part of the OMEGA survey \citep{roman-oliveira19} that are the targets of this study.

The efficiency of the stripping is linearly dependent on the density of the ICM and quadratically on the relative velocity between the galaxy and the environment \citep{gunngott72}. There are two triggering mechanisms, that can act simultaneously, in the stripping of an infalling galaxy: a significant increase in the ICM density (e.g. approaching the centre of a cluster) and/or a high relative velocity between the galaxy and the surrounding medium (e.g. the region between merging clusters). The latter has been thoroughly investigated for the case of Abell 901/2 system in \cite{ruggiero19} where they find regions in the system where ram pressure stripping could be enhanced due to a possible merger between the substructures, explaining the spatial distribution and the large number of candidates of the observed sample of ram pressure stripping candidates. This would confirm previous tentative results that suggest that jellyfish galaxies can be more commonly found in galaxy cluster interactions \citep{mcpartland16, owers12}.

Recent research also finds that although the star formation quenching and the morphological transformation both happen to galaxies as part of their evolution, there is a time delay between these processes \citep{kelkar19, cortese19}. Investigating morphological characteristics of galaxies that are currently going through a major change both in their star formation rates and overall structure can provide insight on whether both changes are linked and how they take place.

So far, very little attention has been paid to the morphological analysis of galaxies with irregular properties, such as jellyfish galaxies.
One study by \cite{mcpartland16} analyses a set of jellyfish galaxies from a morphometric point of view with the main goal of finding a larger sample of ram pressure stripping candidates.
Nonetheless, this analysis can be extremely useful to assess the physical changes that these galaxies are undergoing.

In this paper, we set out to investigate the morphological features of candidate galaxies undergoing ram pressure stripping in a sample of 73 ram pressure stripping candidates in A901/A902 at z$\sim$0.165.
We direct the reader to find more information on the sample in \cite{roman-oliveira19}, where we describe the selection and its main properties, and in \cite{ruggiero19} that further explores the origin of the possible ram pressure stripping events.
Our goal is to understand how the ram pressure stripping mechanism is modifying their structure and its contribution to the scenario of quenching and morphological evolution in dense environments.
%We also investigate the role that jellyfish galaxies can play in probing the multi-cluster dynamics through their line of sight and projected velocities.
We perform the morphometric analysis using the \textsc{morfometryka} algorithm \citep{ferrari15} to measure trail vectors, surface brightness profiles and other morphometric quantities.

%Structure of the paper
This work is organised as follows: in Section~\ref{sec:data} we detail the data, sample and methods used; in Section~\ref{sec:results} we show the results of the morphometric analysis for trail vectors, surface brightness profiles and curvature; and in Section~\ref{sec:conclusion} we summarise our conclusions.
%obs: which cosmology is used, scale images, etc.
We adopt a $\text{H}_0 = 70\text{kms}^{-1}\text{Mpc}^{-1}, \Omega_{\Delta} = 0.7$ and $\Omega_{M}=0.3$ cosmology through this study.

\section{Data and Methods}\label{sec:data}%ok

\subsection{Abell 901/2}
Abell 901/2 is a multi-cluster system at z$\sim$0.165 composed of four main sub-cluster structures and filaments. It has been intensely studied by the STAGES collaboration \citep{gray09} and, more recently, by the OMEGA survey \citep{roman-oliveira19, wolf18, weinzirl17, rodriguezdelpino17, chies-santos15} in many different wavelengths. It is a particularly interesting system because of its large galaxy population and diverse environments, making it suitable for detailed studies of galaxy evolution through a vast range of stellar masses and environments.

\subsubsection{Sample}
In this study we make use of Hubble Space Telescope (HST) observations in the ACS/F606W passband of the Abell 901/2 multi-cluster system where a sample of 73 ram pressure stripping candidates has been previously selected through visual inspection as part of the OMEGA survey.
Along with the HST imaging, we use a model PSF (point spread function) obtained with Tiny Tim \citep{krist93}.

Although the jellyfish galaxy tails are not as visible in optical bands as in X-rays or H$\alpha$, the stellar disc shows a disturbed morphology that can be evidence of more extreme disturbances in other wavelengths \citep{poggianti19b}. This can be used to select samples of ram pressure stripping candidates, like the ones used in this paper. Searching for ram pressure stripping features on optical images is an efficient and economic method of finding ram pressure stripping candidates that has been employed on many works through visual inspection \citep{owers12, rawle14, ebeling14, poggianti16}. The disturbed morphologies of these candidates can be due to ram pressure stripping, however samples selected this way also have some degree of contamination by minor mergers or tidal interactions. Therefore, only follow-up studies in other passbands could rightly confirm the origin of the stellar disc disturbance.

%A morphometric study of these ram pressure stripping candidates is useful to understand what disturbances in the stellar disc are being favoured in the visual selection. establishing a galaxy morphometry system is that we can seek structures, in the quantitative morphology parameter space, that may yield clues for the physical reasons for their formation and evolution that are not visible in the currently human-based mode.

The F606W passband has an effective wavelength midpoint ($\lambda_{eff}$) around 5777\AA; at $z\sim0.165$ we are thus covering the rest-frame R-band around 6730\AA.
This interval covers intermediate/old stellar populations that contribute to the continuum emission in this red part of the spectrum and to some extent young stellar populations by encompassing the H$\alpha$ emission. In this range of wavelengths, the presence of dust can significantly obscure star formation in nearly edge-on galaxies \citep{wolf18}, which composes a minority of the sample. Besides, the morphometric measurements of the stellar disc should be mostly unaffected.

%As for the presence of dust, \citet{wolf18} finds that the star formation rates measured in H$\alpha$ are underestimated by a factor of 2 in nearly edge-on galaxies.

The sample was selected in \cite{roman-oliveira19} and the selection method was conducted mirroring the works of \cite{poggianti16} and \cite{ebeling14}.
This is the largest sample up to date for a single system containing galaxies with morphological signatures linked to ram pressure stripping effects, such as tails and bright knots of star formation.
The galaxies are selected in different categories according to the prominence of the ram pressure stripping features in their morphologies. The strongest candidates are grouped in JClass 5, the weakest candidates in this sample are grouped in JClass 3 and the intermediate candidates are grouped in JClass 4.
For further details on the the selection and eligibility criteria and basic physical properties of the sample refer to \cite{roman-oliveira19}.

\begin{figure*}
    \centering
    \includegraphics[]{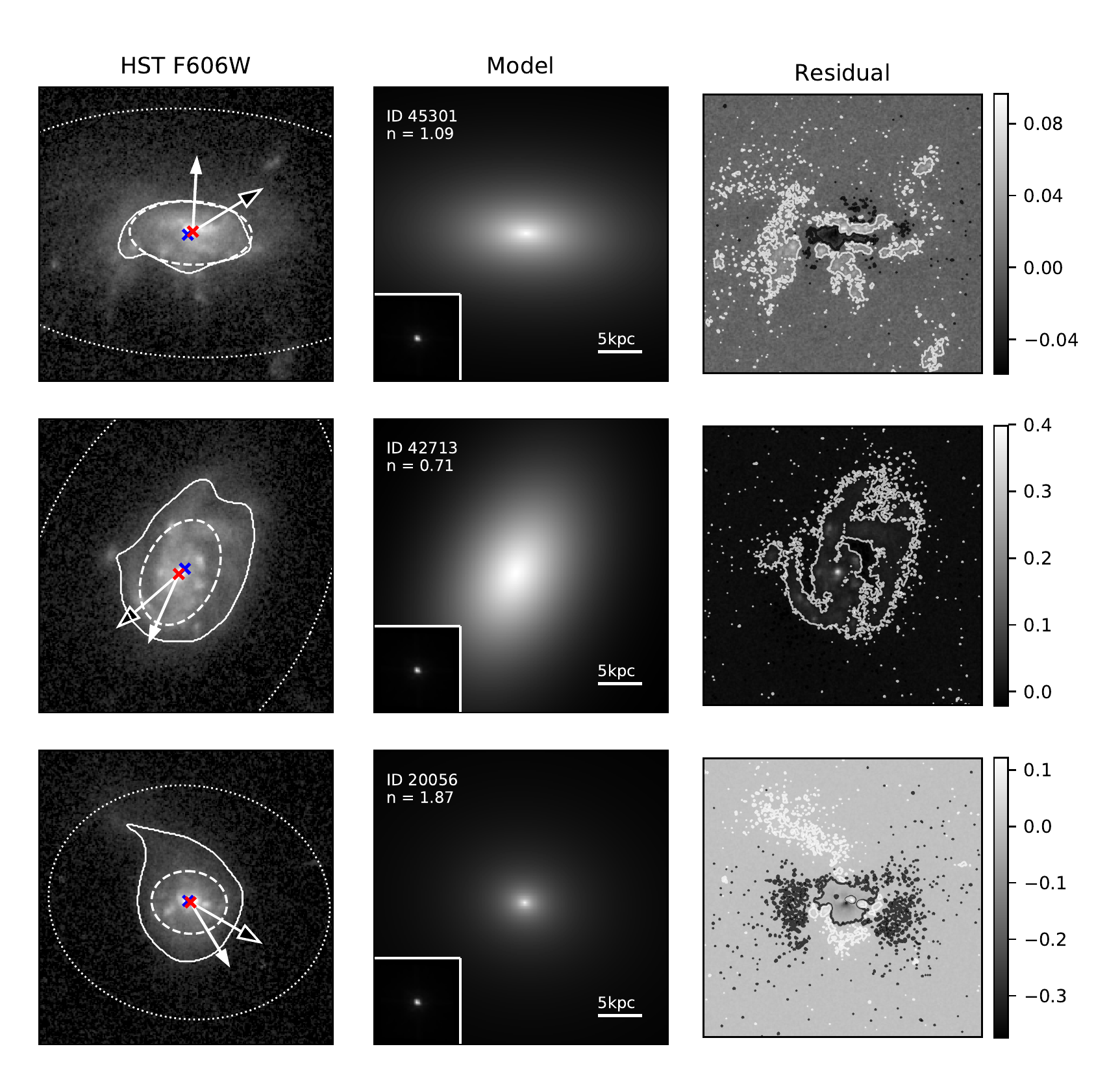}
    \caption{\textsc{morfometryka} analysis of three galaxies with the strongest ram pressure stripping features in A901/A902. Left column: original HST image. The outer dotted ellipse represents twice the Petrosian region, the dashed inner ellipse represents twice the effective radius of the S\'ersic model. The solid line is the segmented region. The black headed arrow shows the morphometric trail vector and the white headed arrow shows the visually assigned trail vector. The peak of light is represented by a blue cross and the centre of light is represented by a red cross. Middle column: two-dimensional S\'ersic model. The bottom-left square shows the HST/F606W PSF modelled with Tiny Tim. The galaxy ID and S\'ersic index are noted in the top-left corner. Right column: Residual image with its respective colourbar. The contours show regions that have values 3 $\sigma$ above the sky background, for negative value the contours are represented in black and for positive values the contours are represented in white.}
\label{fig:mfmtk}
\end{figure*}

\subsection{Morphometric analysis}
Several techniques have been developed to quantify the physical structures of galaxies in measurable ways. One example is the CASGM non-parametric system that measures concentration, asymmetry, smoothness, Gini coefficient and M20 parameters \citep{lotz04, conselice00, abraham94}.

Our work is based on the \textsc{morfometryka} algorithm that establishes a new method dedicated to morphology classification from a physical standpoint. It includes the parameters cited above as well as entropy (H) and spirality ($\sigma_{\psi}$) as new parameters \citep{ferrari15}.
The most recent version of \textsc{morfometryka} also provides the curvature of the brightness profile with \textsc{kurvature} \citep{lucatelli19}, which is a powerful tool for probing the presence of multiple components in galaxies.
An example of the performance of \textsc{morfometryka} for one of our galaxies displaying signatures of ongoing ram pressure stripping can be seen in Figure~\ref{fig:mfmtk}.

%%%%%%%%%%%%%%%%%%%%%%%%%%%%%%%%%%%%%%%%%%%%%%%%%%%%%%%%%%%%%%%
%%%%%%%%%%%%%%%%%%%%%%%%%%%%%%%%%%%%%%%%%%%%%%%%%%%%%%%%%%%%%%%
%%%%%%%%%%%%%%%%%%%%%%%%%%%%%%%%%%%%%%%%%%%%%%%%%%%%%%%%%%%%%%%

\section{Results}\label{sec:results}

\subsection{Morphometric Trail Vectors}%ok

%definition and asymmetry proxy
\subsubsection{Definition and use as an asymmetry measurement}%ok

Within \textsc{morfometryka}, we implement an automatic way to define the direction of motion. As jellyfish galaxies fall into the galaxy cluster they leave a trail of material behind. This trail hints at the projected motion around the system. This method has been adopted so far mainly through visual inspection in a number of works \citep{roman-oliveira19, ebeling14, smith10}, but most recently \cite{yun19} measured trail vectors for 800 ram pressure stripping candidates in the Illustris TNG by defining the direction of a vector between the density-weighted mean to the galaxy centre positions.
Here, we perform something similar to \cite{yun19} from the standpoint of observations in which we measure a trail vector ($\textbf{x}$) from the position of the centre of light to the peak of light. The peak of light is correlated to the centre of the galaxy and should remain the same before and after undergoing ram pressure stripping. The centre of light is a density-weighted mean of the light distribution that is highly affected by perturbations in the morphology. We measure the morphometric trail vector with \textsc{morfometryka} following: $\textbf{x} = (x_0, y_0)_{\rm peak} - (x_0, y_0)_{\rm CoL}$. For more details on how these components are measured we refer the reader to \citet{ferrari15}.

\begin{figure}
    \centering
    \includegraphics{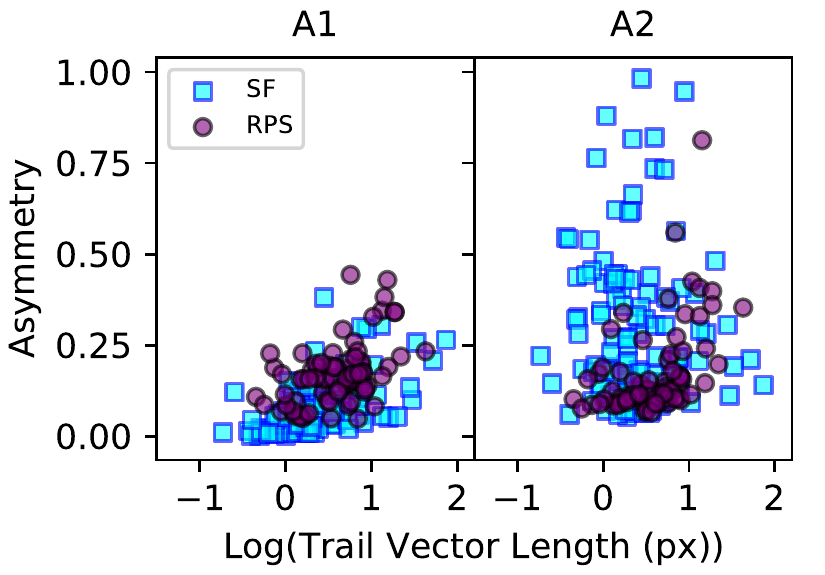}
    \caption{Asymmetry versus trail vector length in ram pressure stripping candidates and star forming galaxies in A901/A902. In the left panel we show the asymmetry parameter A1, defined in \citet{abraham96}, and in the right panel we show the parameter A2, defined in \citet{ferrari15} that is less sensitive to the sky.}
    \label{fig:tvl}
\end{figure}

Not only does this method give a quantifiable measurement of the orientation of the projected motion of the galaxies, it is also more sensitive to slight perturbations in the structure that visual inspection cannot account for. The offset between the two points can also be considered a proxy for asymmetry, since the peak of light and centre of light coincide in an axisymmetric structure with a surface brightness profile that decays with increasing radius, such as a pure disc component. In Figure~\ref{fig:tvl} we show a comparison between the trail vector length (TVL) and two morphometric asymmetry parameters, A1 and A2. We test this both for the trail vectors measured for the ram pressure stripping candidates and for a control sample of star forming galaxies that do not show morphological features of ram pressure stripping as for our selection. The galaxies that form this control sample were selected as star forming galaxies based on their H$\alpha$ emission as detailed in \cite{rodriguezdelpino17}. A1 and A2 are parameters determined by the summation of the residual of an image with its rotated counterpart. A1 is measured as defined by \citet{abraham96}, by subtracting the rotated galaxy image ($I_{\pi}$) from the original galaxy image ($I$) within the Petrosian radius and without subtracting the sky, following:
\begin{equation}
A_1 = \frac{\text{abs}(I - I_{\pi})}{2I}  
\end{equation}
While A2 is measured as defined in \citet{ferrari15} and uses a Pearson correlation coefficient ($r()$) to avoid contamination from the sky, following:
\begin{equation}
A_2 = 1 - r(I,I_{\pi})
\end{equation}
The main difference between A1 and A2 is that A1 is sensitive to the sky background while A2 is unaffected by it. We measure a Pearson correlation for the TVL with A1 and A2. In Table~\ref{tab:tvl} we show the resulting Pearson coefficients and respective p-values. We find that for the ram pressure stripping candidates they are related with great certainty (p-values of 2e-05 and 2e-06). However, for the other star forming galaxies we find a correlation of the TVL with A1, but no correlation between TVL and A2. Many star forming galaxies have low A1 values and high A2 values. This can be due to the fact that although A2 is unaffected by the sky background, it tends to be more sensitive than A1 to small perturbations inside a galaxy, for example spiral arms or a clumpy disk. Therefore, galaxies that do not have a very asymmetric morphology, but have this perturbations will not follow a correlation with TVL. Another important scenario is that some galaxies may have large A1 or A2 values but not be unilaterally asymmetric, in which case the TVL will be relatively small for the asymmetry parameters measured, breaking up the correlation between each other. The correlation between TVL and both the asymmetry parameters probed suggests that the morphometric trail vectors are a good parameter for measuring unilateral asymmetries.
This method vectors can be applied to large datasets and aid the analysis and identification of new ram pressure stripping candidates, which is a large improvement over visually assigned trail vectors.

\begin{table}
    \centering
\begin{tabular}{lcc}
\hline
\hline
       & Pearson coefficient   &  p-value \\
      \hline
RPS, TVL and A1    & 0.478  & 2e-05  \\
RPS, TVL and A2    & 0.505  & 5e-06  \\
\hline
SF, TVL and A1     & 0.564 & 3e-11  \\
SF, TVL and A2     & -0.05  & 0.6  \\
\hline
\hline
\end{tabular}
\caption{Statistics of the Pearson correlation between trail vector length (TVL) and asymmetry parameters A1 and A2 for ram pressures stripping candidates (RPS) and star forming galaxies (SF).}
\label{tab:tvl}
\end{table}

%comparison and spatial distribution

\subsubsection{Comparison to the visually assigned trail vectors}%ok

\begin{figure*}
    \centering
    \includegraphics[]{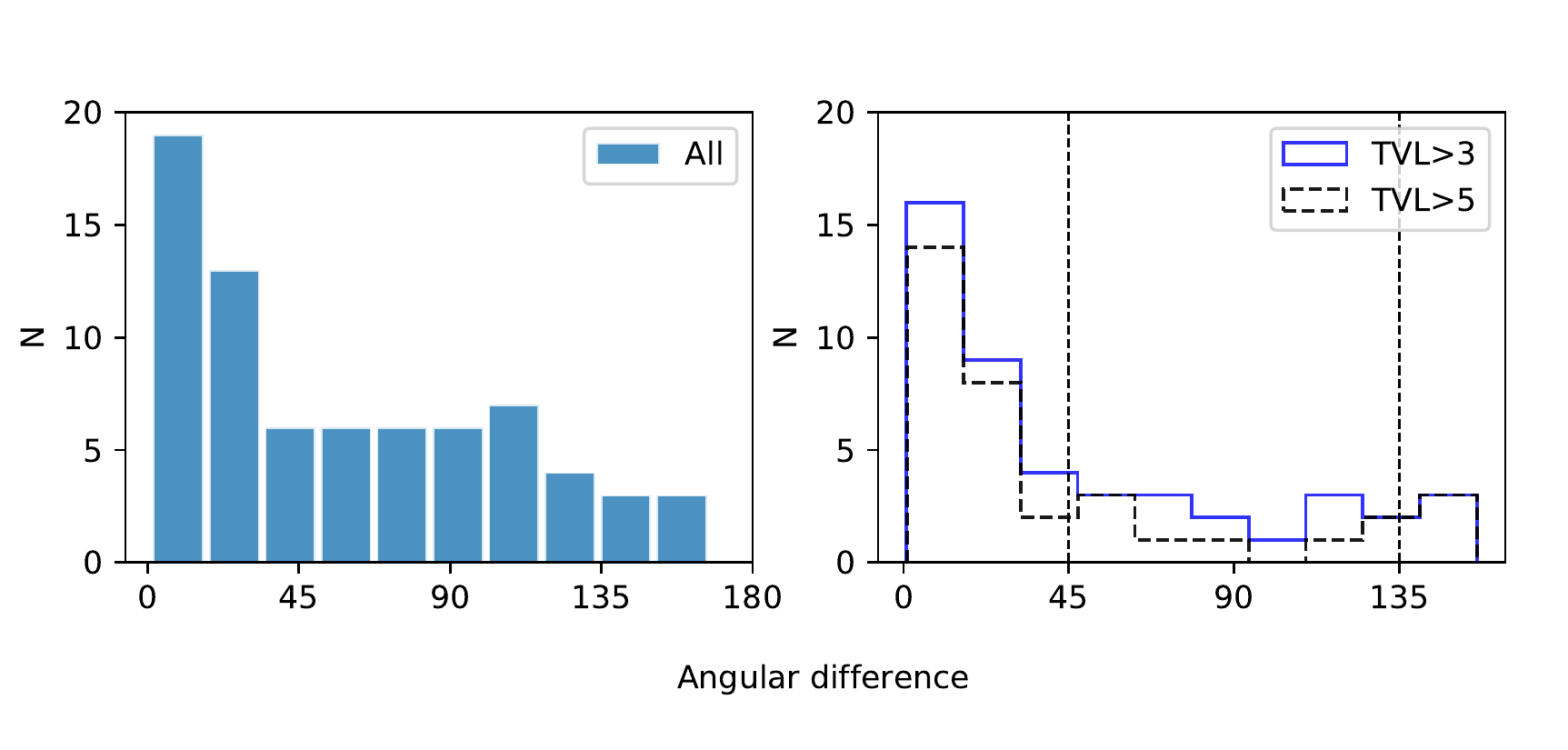}
    \caption{Histogram of the angular difference between morphometric and visually assigned trail vectors. Left panel: for all the ram pressure stripping candidates. Right panel: for ram pressure stripping candidates with a trail vector length of at least 3px and 5px ($\sim$0.4kpc and $\sim$0.6kpc at z$\sim$0.165, respectively). The vertical dashed lines mark angular differences of 45 and 135 degrees.}
    \label{fig:tvhist}
\end{figure*}

\begin{figure*}
    \centering
    \includegraphics[]{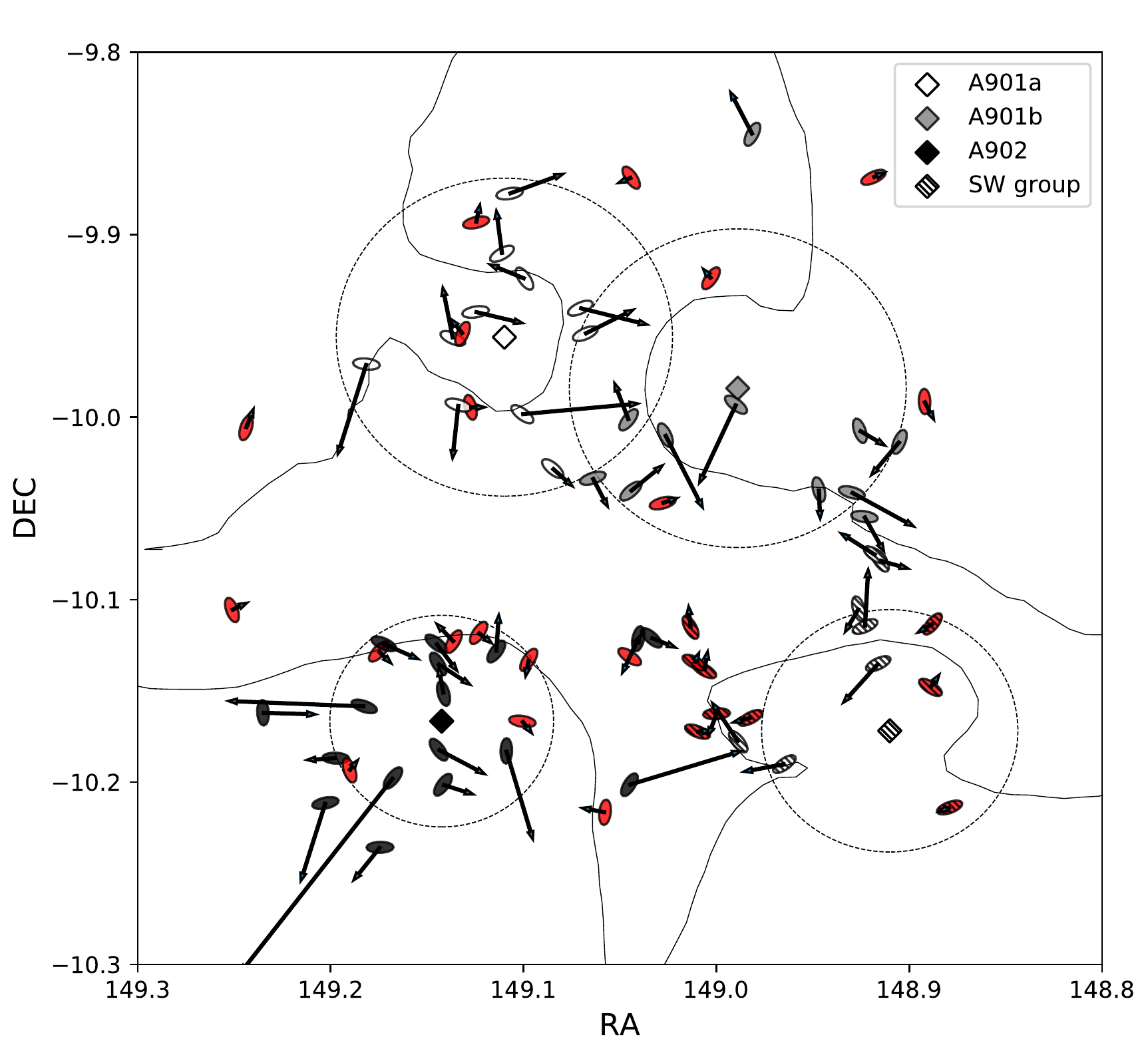}
    \caption{The spatial distribution of the ram pressure stripping candidates and their morphometric trail vectors tracing their projected motion on the sky. The centres of each subcluster is marked with a diamond symbol according to the legend. The ram pressure stripping candidates are represented with ellipses with the measured position angles and their colours match the subcentre that they are closest to in projected distance. The red coloured markers identify the galaxies with TVL smaller than 3px. The arrows represent the measured trail vectors and the length is proportional to the distance between the centre and the peak of the light distribution. The continuous lines show the expected region where ram pressure stripping would be triggered in response to the merging clusters as detailed in \citet{ruggiero19}. The dotted circles represent the virial radius (R$_{200}$) of each subcluster used in \citet{ruggiero19}.}
    \label{fig:spatial}
\end{figure*}

In Figure~\ref{fig:tvhist} we compare the visually assigned trail vectors from \citet{roman-oliveira19} with the morphometric trail vectors presented in this work by calculating the angular difference between both vectors. We are considering 45 degrees as the threshold to which we consider as a good agreement between the vectors since it would still point towards the same general direction and it is comparable to the disagreement between the vectors suggested by different inspectors during the visual assignment. Similarly, we consider an angular difference of 135 degrees or more to be a good agreement in direction although it is suggesting an opposite pointing. We find that about half of the galaxies can be considered in good agreement by these standards. However, if we restrict this comparison to only the galaxies that have a TVL of at least 5 pixels, which at z$\sim$0.165 is around $0.6kpc$, about three quarters of the galaxies considered are in good agreement. This suggests that the direction of the morphometric trail vectors are more reliable for higher TVLs and should be considered carefully for galaxies with less prominent morphological disturbances. In Figure~\ref{fig:spatial} we show the spatial distribution of the ram pressure stripping candidates with the new morphometric trail vectors. Similarly to what was found with the visually assigned trail vectors, we see no correlation between the direction of the projected motion of the candidates in the system.

%How robust is the new, morphology-based algorithm and what are its potential systematic biases?
Besides, there is an intrinsic bias on measuring the coordinates of the peak of light, in the case of galaxies that do not have a definitive centre or when the peak is found in a bright star forming region outside the centre.
As for measuring the centre of light, the coordinates are most sensitive to the shape selected to represent the morphology of the galaxy. In the case of \textsc{morfometryka} we are calculating the centre of light inside the segmented region -- this region is shown in Figure~\ref{fig:mfmtk}. The \textsc{morfometryka} segmentation selects a region that has a significant intensity above the background sky -- the region is selected through applying histogram thresholding on a filtered image to avoid sharp edges (see \citet{ferrari15} for more details). This segmentation is sensitive to the size of the image analysed, which is why it is important to have an image stamp large enough to cover the structures of interest, but small enough that it will not introduce contamination from nearby sources.

Lastly, besides the scenarios we commented, in some cases, the disagreement between the morphometric and the visually assigned trail vectors can be due to the morphometry being more sensitive to disturbances that are too small for the inspectors to correctly assign a vector, in which case the morphometric trail vector is superior to the visually assigned one. We emphasise that this method has its limitations regarding projection effects and it works best for edge-on/inclined galaxies. In the case of face-on galaxies it may still be able to provide an accurate orientation of the trail vector but it might underestimate the TVL. This discrepancy can be better seen in \citet{yun19}, where trail vectors were estimated in a similar way for jellyfish galaxies in the Illustris TNG simulations. Some galaxies show clear extended jellyfish tails when seen edge-on, but do not look as disturbed when face-on. Additionally, the reader can visualise all the results in Appendix~\ref{ap:atlas}, where we show the galaxy stamps, segmented areas, morphometric and visually assigned trail vectors.

%%%%%%%%%%%%%%%%%%%%%%%%%%%%%%%%%%%%%%%%%%%%%%%%%%%%%%%%%
%%%%%%%%%%%%%%%%%%%%%%%%%%%%%%%%%%%%%%%%%%%%%%%%%%%%%%%%%
%%%%%%%%%%%%%%%%%%%%%%%%%%%%%%%%%%%%%%%%%%%%%%%%%%%%%%%%%

\subsection{Surface Brightness Profiles}

%morfometryka results
\subsubsection{\textsc{morfometryka} and S\'ersic indices distribution}%ok

With \textsc{morfometryka}, we model the surface brightness profiles of the ram pressure stripping candidates with a single two-dimensional S\'ersic Law \citep{sersic68} to investigate the light distribution properties.
It is important to note that this does not model the distorted tails, but it does give an overall assessment of the light concentration in the galaxies.
In Figure~\ref{fig:mfmtk}, we showcase the \textsc{morfometryka} models and residuals for three example galaxies with S\'ersic indices that represent three groups of surface brightness: disc-like (n$\sim$1), more concentrated than a disc (n$>$1) and less concentrated than a disc (n$<$1). The galaxies chosen (IDs 45301, 42713 and 20056) were classified in \cite{roman-oliveira19} as JClass 5, which means they have the strongest features of ram pressure stripping among the sample.

%sersic distribution
We first analyse the distribution of S\'ersic indices of the modelled profiles of the ram pressure stripping candidates and compare it to the other star forming galaxies in the system. In this, we find that the S\'ersic indices distribution for the ram pressure stripping is centred around n$\sim$1, with a median $\tilde{\text{n}}=1.06$. We account also for the dependency of stellar mass with S\'ersic index by considering two separate bins of mass below and above M$_{*}$ = 10$^{9.5}$ M$_{\odot}$. We chose this threshold as it lies in between the median mass of both the candidates and the control sample. In Table~\ref{tab:ndist} we show the parameters measured for the distribution of S\'ersic indices for both samples and bins. Both distributions are similar, the main difference seems to be that the ram pressure stripping candidates are more tightly distributed around the mean and that the division in stellar mass bins does not affect the distribution.

\begin{table}
    \centering
\begin{tabular}{lcccc}
\hline
\hline
       & N   & $\tilde{\text{n}}$ & $\overline{\text{n}}$ & $\sigma_{\text{n}}$  \\
      \hline
RPS    & 73  & 1.06   & 1.18 & 0.61 \\
RPS$_{\text{low mass}}$ & 22  & 1.03   & 1.04 & 0.35 \\
RPS$_{\text{high mass}}$ & 51  & 1.08   & 1.24 & 0.69 \\
\hline
SF     & 112 & 1.03   & 1.48 & 2.14 \\
SF$_{\text{low mass}}$  & 89  & 0.98   & 1.51 & 2.37 \\
SF$_{\text{high mass}}$   & 23  & 1.12   & 1.34 & 0.73 \\
\hline
\hline
\end{tabular}
\caption{Distribution of S\'ersic indices for the ram pressure stripping candidates (RPS) and star forming galaxies (SF) in A901/A902. We show the values for the full samples and for bins of stellar mass above and below M$_{*}$ = 10$^{9.5}$ M$_{\odot}$. The columns show the number of galaxies (N), median ($\tilde{\text{n}}$), mean ($\overline{\text{n}}$) and stardard deviation ($\sigma_{\text{n}}$) of the S\'ersic indices.}
\label{tab:ndist}
\end{table}

\subsubsection{\textsc{ellipse} and surface brightness curvature profiles}

\begin{figure*}
    \centering
    \includegraphics[]{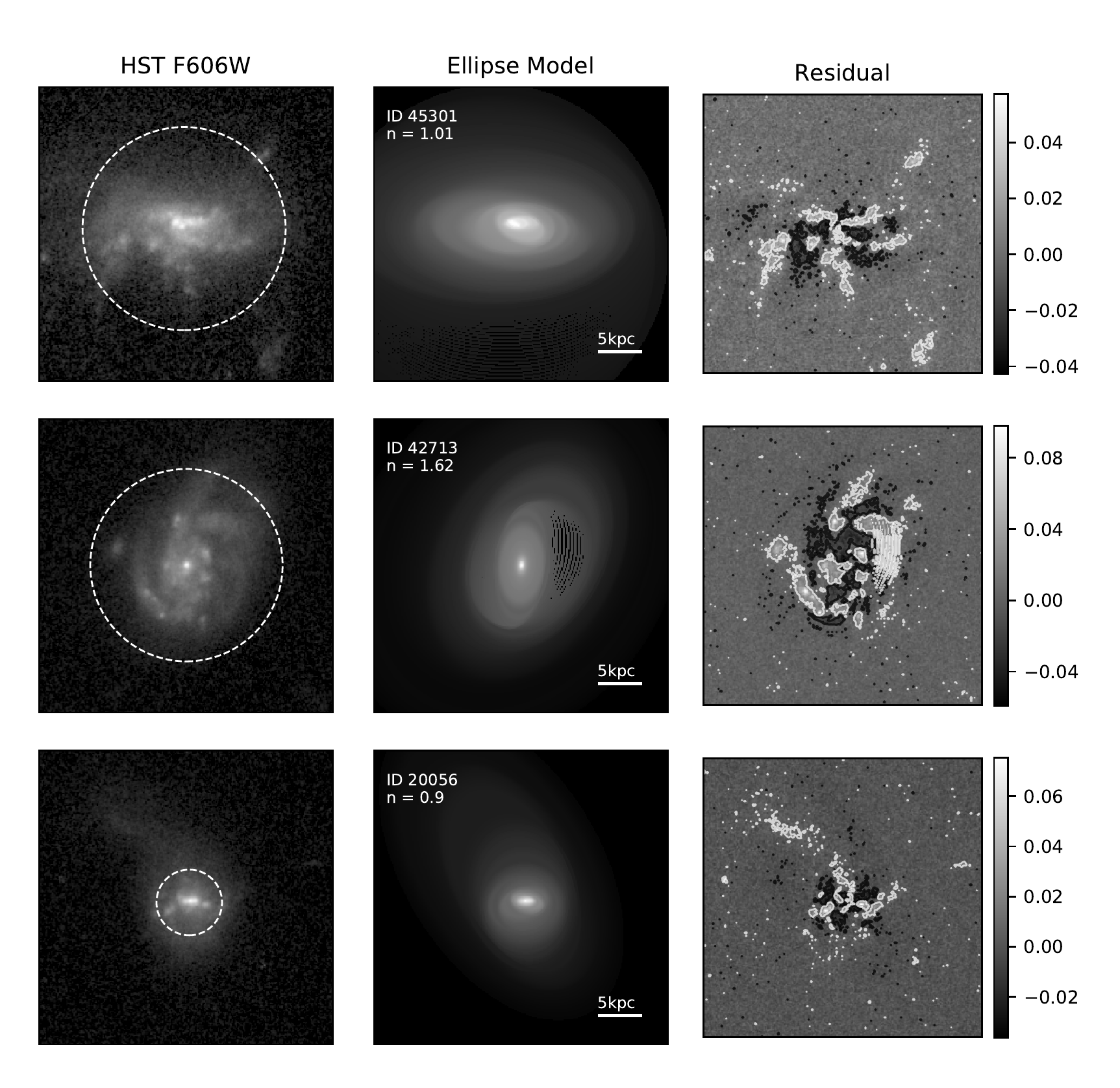}
    \caption{\textsc{ellipse} analysis of the same three galaxies from Figure~\ref{fig:mfmtk}. Left column: original HST image. The dashed circle represents twice the effective radius of the model. Middle column: \textsc{ellipse} model. The ID and S\'ersic index fitted are noted in the top-left corner. Right column: residual image and its respective colourbar. The contours and contrast are the same as those in Figure~\ref{fig:mfmtk}.}
    \label{fig:ellipse}
\end{figure*}

We find from the S\'ersic distribution that the overall surface brightness profile of the ram pressure stripping candidates can be approximated by discs. However, \textsc{morfometryka} cannot fit most of the details that stem from the irregular structure. To further investigate the light distribution of the sample, we use the \textsc{iraf/ellipse} task \citep{jedrzejewski87}. \textsc{ellipse} achieves a more accurate measurement of the brightness profile by fitting several ellipses of increasing semi-major axes and different position angles, being more sensitive to irregular structures of galaxies.
In Figure~\ref{fig:ellipse} we show the results from \textsc{ellipse} for the same galaxies we analyse in Figure~\ref{fig:mfmtk}. We maintain the same contrast used in the previous figure to allow the reader to visually compare the results obtained from the two algorithms.

We assess the quality of both models by evaluating the residuals from the contours shown in the right panels of Figure~\ref{fig:mfmtk} and Figure~\ref{fig:ellipse}. The contours highlight the regions 3 standard deviations below (black contours) or above (white contours) the sky background. Therefore, the black contours show regions that are being overfitted by the model and the white contours show clumpy star forming regions, arms or irregular structures not represented in the model. We calculate ratios of residual to the original image and we found that neither codes tend to overfit the galaxies, as the ratios of the black contoured regions to the original are around 0.03 for all galaxies -- except for \textsc{morfometryka} fitting the galaxy 20056 with a ratio of 0.2. As for white contours, \textsc{ellipse} has a much better performance with ratios or residual to original of 0.06 for all three galaxies, effectively covering most of the emission of the galaxy even for the irregular components. In that aspect, \textsc{morfometryka} ranges in ratios of 0.08 (ID 20056), 0.14 (ID 45301) to 0.35 (ID 42713).

\begin{figure*}
    \centering
    \includegraphics{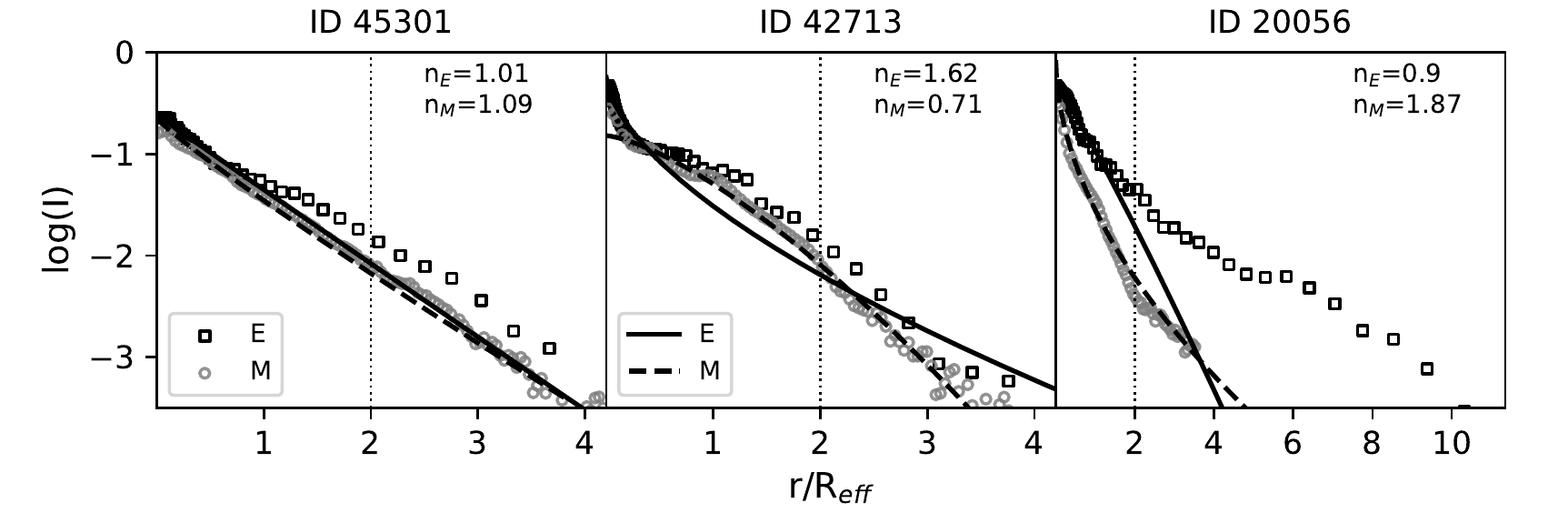}
    \caption{Surface brightness profiles for the three galaxies show in Figure~\ref{fig:mfmtk} and Figure~\ref{fig:ellipse}. The vertical dotted lines mark twice the effective radius, this relates to the dashed circles in Figure~\ref{fig:ellipse}. The black square and gray circle markers show the \textsc{ellipse} (E) and \textsc{morfometryka} (M) surface brightness profile, respectively. The black solid and dashed lines are the best single S\'ersic fit for \textsc{ellipse} and \text{morfometryka}, respectively. The S\'ersic indices for both fits are noted in each panel.}
    \label{fig:bp}
\end{figure*}

In Figure~\ref{fig:bp} we show the surface brightness profiles measured with \textsc{morfometryka} and \textsc{ellipse} for the three galaxies as well as the best fit S\'ersic models. In all the three cases we see a large scale structure that has a concave shape in the surface brightness profile. However, even though \textsc{ellipse} retrieves the light distribution of the galaxy, a single S\'ersic fit does not represent well all the features we see in the surface brightness profile, this is especially true for the case of ID 20056 that has extended emission in comparison to its effective radius. These galaxies seem to have multiple structural components and a single S\'ersic fit can only fit one of these components. In the case of ID 20056, the S\'ersic fit best represents the inner region, but not the more extended concave profile. A similar situation occurs for the other two galaxies in both \textsc{morfometryka} and \textsc{ellipse} measured profiles. For evaluating these structures we take advantage of the tool \textsc{kurvature} \citep{lucatelli19} which measures the curvature of a surface brightness profiles by calculating its concavity. The concave shapes we find are related to a negative curvature which is related to low concentrated of light in surface brightness profiles, such as S\'ersic fits with n < 1.

\begin{figure}
    \centering
    \includegraphics[]{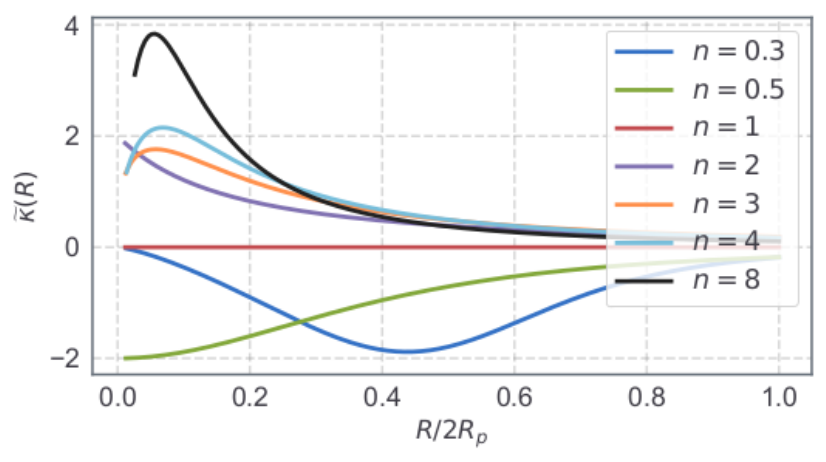}
    \caption{Curvature ($\tilde{k}(R)$) for S\'ersic profiles of different S\'ersic indices. Negative curvature profiles are associated with structures of low S\'ersic indices, a null curvature profile represents a pure disc and positive curvature profiles are associated with high S\'ersic indices that follow light distributions more concentrated than a pure disc.}
    \label{fig:kur_n}
\end{figure}

To better understand curvature measurements, in Figure~\ref{fig:kur_n} we show the relation between S\'ersic indices and the curvature of surface brightness profiles. S\'ersic profiles with high S\'ersic indices (n $>$ 1) have positive curvature profiles, while low S\'ersic indices (n $<$1) have negative curvature profiles and pure discs (n=0) have null curvature. This is a powerful tool to assess the concentration of light distribution and discriminate between different structural components in a galaxy without depending on a parametric model. Therefore, a negative curvature profile is directly related to a region of low concentration of light in the surface brightness profile, the area of the curvature profile also correlates with S\'ersic index. It is important to note that the negative areas are unlikely to be due to noise. Curvature measurement is sensitive to transitions between two regions with different brightness profiles. Hence, the transition between a decreasing brightness profile of a galaxy meeting the constant background noise would be interpreted by \textsc{kurvature} with a positive curvature. In the cases where the curvature diverges in outer regions, most are in the positive direction. Following this same reasoning, concave regions could be associated with regions that lack light in respect to their surroundings, such as in ring or bar structures. Perhaps regions with high dust extinctions can also contribute to the phenomenon. However, the ram pressure stripping candidates we are probing do not necessarily contain more dust than the star forming galaxies in the control sample, therefore, the presence of dust affects both samples in similar ways.

\begin{figure*}
    \centering
    \includegraphics[]{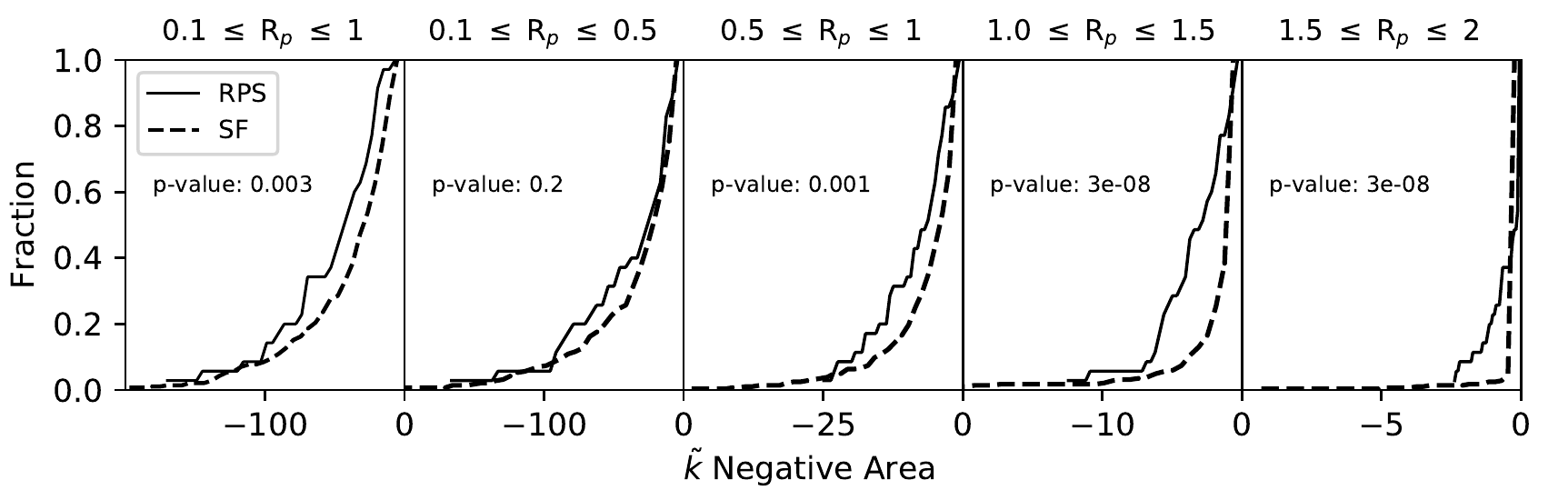}
    \caption{Cumulative distribution of the total negative area in the curvature profiles, within a given radius, measured with \textsc{morfometryka} tool \textsc{kurvature} for the \textsc{ellipse} brightness profiles for JClass 4 and 5 ram pressure stripping candidates (N=35) as a solid line and star forming galaxies as a dashed line in the A901/A902 system. The left panel accounts the surface brightness up to 2 R$_p$, the following panels are divided into radial bins of 0.5 R$_p$. The p-values shown are calculated with a KS test.}
    \label{fig:kurv_negarea}
\end{figure*}

We quantify the presence of concave features by measuring the cumulative negative area in the surface brightness profiles of the ram pressure stripping candidates and the control sample of star forming galaxies. To avoid contamination from galaxies with weak signatures of ram pressure stripping, we are considering only the JClass 4 and JClass 5 ram pressure stripping candidates (N=35). We show the cumulative histograms in Figure~\ref{fig:kurv_negarea} where we compare both groups of galaxies with a KS test across 2 Petrosian radii (R$_p$) and in four different radial bins. We neglect the central values in r $\leq$ 0.1 R$_p$ due to the curvature profile being unstable in the inner regions. We find that both samples are significantly different when looking at the full radius range and the outer radial bins (r $\geq$ 0.5 R$_p$), with the ram pressure stripping candidates always having more negative area than the star forming galaxies. This is more prominent for the radial bins above 1.0 R$_p$. In this plot we consider all galaxies JClass 4 and 5 regardless of their TVL, however we perform the same test considering only the JClass 4 and 5 galaxies with TVL>3px and with TVL>5px and the results are unaffected.These results suggest that ram pressure stripping may systematically alter the galaxy morphology by broadening the surface brightness profiles effectively creating galaxies that have the stellar component less concentrated than a pure disc in the outer regions (0.5 R$_p$ $\geq$ r $\leq$ 2.0 R$_p$).

Concave features in surface brightness profiles are not unique to the ram pressure stripping candidates analysed here, but seem to be present more often in our ram pressure stripping candidates than in normal star forming galaxies. These concave features, when seen in normal disc galaxies, are usually associated with structural components such as rings or bars, which are not prominent in our sample. However, these features can also be associated with an overall low concentrated light distribution, such as seen in the surface brightness profiles of some dwarf galaxies \citep{ludwig12} or ultra diffuse galaxies \citep{liao19}.

\section{Conclusions}\label{sec:conclusion}%ok

Following the studies on the star formation rates and spatial distribution of the ram pressure stripping candidates at the A901/A902 multi-cluster system \citep{roman-oliveira19, ruggiero19}, we attempt to use their morphological structure as a probe to expand our understanding of their evolution.
We perform a morphometric analysis using the \textsc{morfometryka} algorithm \citep{ferrari15} and the \textsc{iraf} task \textsc{ellipse} \citep{jedrzejewski87} for independent surface brightness profiles measurements. Our two main results are:

%Speculate about results
\begin{itemize}
    \item We define a robust morphometric method for measuring trail vectors in jellyfish galaxies based on the spatial difference between the peak and centre of the light distribution in galaxies. This can also be used as a proxy of morphological asymmetry.
    
    \item  Our analysis of the surface brightness profiles finds a significant presence of low concentration regions that can be seen as concavities in the surface brightness profiles, we quantify these regions by measuring the curvature \citep{lucatelli19}. When these are compared to the normal star forming galaxies in the same system, the ram pressure candidates show larger areas of negative curvature in the outer regions of their surface brightness profiles. This suggests that the extreme ram pressure that produces jellyfish features also serves to broaden the surface brightness profiles creating regions that are less concentrated than pure discs.

\end{itemize}

%State the contributions/importance
The findings reported here shed new light on the possible next steps in the morphological evolution of galaxies undergoing ram pressure stripping in dense environments. We suggest that, at least temporarily, extreme events of ram pressure stripping may affect the morphology by broadening the surface brightness profiles of galaxies. Additionally, the implementation of morphometric trail vectors is an important step towards systematic selection and analysis of projected motions of new ram pressure stripping candidates, as well as another useful tool to quantify asymmetry.

These are preliminary findings on the morphological transformation of ram pressure stripping candidates. The details on how ram pressure stripping could alter the morphology of the stellar disc are still largely unknown. A further investigation of the morphometric properties of these galaxies in a different passband can retrieve information on how the morphology of different physical tracers is being affected. Particularly, applying the same morphometric analysis on the OMEGA H$\alpha$ emission and building H$\alpha$ morphology profiles \citet{koopmann04} can unveil the extent and concentration of the star formation spatially, whether it is being enhanced or suppressed in different regions of the galaxies and if it is related to the concave regions we see in the F606W passband.

\section*{Acknowledgements}%ok
This work was conducted as part of an M.Sc. thesis in a Federal University despite the hard current policies of Brazil's far-right government against education and science.
We thank the referee Harald Ebeling for the constructive comments which substantially helped to improve the quality of this work. We also thank Alfonso Arag\'on-Salamanca for comments in the original manuscript.
This study was financed in part by the \textit{Coordena\c{c}ao de Aperfei\c{c}oamento de Pessoal de N\'ivel Superior} - Brasil (CAPES) Finance Code 001, the \textit{Programa de P\'os Gradua\c{c}\~{a}o em F\'isica} of \textit{Universidade Federal do Rio Grande do Sul} and PROPESQ/UFRGS.
ACS acknowledges funding from the Brazilian agencies \textit {Conselho Nacional de Desenvolvimento Cient\'ifico e Tecnol\'ogico} (CNPq) and the \textit{Funda\c{c}\~{a}o de Amparo \`a Pesquisa do Estado do Rio Grande do Sul} (FAPERGS) through grants PIBIC-CNPq, CNPq-403580/2016-1, CNPq-310845/2015-7, PqG/FAPERGS-17/2551-0001, PROBIC/FAPERGS and L'Or\'eal UNESCO ABC \textit{Para Mulheres na Ci\^encia}.
BRP acknowledges support from the Spanish Ministerio de Econom\'{i}a y Competitividad through the grant ESP2017-83197-P.
STSDAS and PyRAF are products of the Space Telescope Science Institute, which is operated by AURA for NASA.

\section*{DATA AVAILABILITY}
The data underlying this article will be shared on reasonable request to the authors and the PIs of STAGES and OMEGA.

%%%%%%%%%%%%%%%%%%%%%%%%%%%%%%%%%%%%%%%%%%%%%%%%%%



\bibliographystyle{mnras}
\bibliography{refer}

%%%%%%%%%%%%%%%%%%%%%%%%%%%%%%%%%%%%%%%%%%%%%%%%%%

%%%%%%%%%%%%%%%%% APPENDICES %%%%%%%%%%%%%%%%%%%%%

\appendix

\section{ATLAS}\label{ap:atlas}

In this section we show the HST F606W stamps for the ram pressure stripping candidates with the segmented region shown as white contours, the morphometric trail vectors shown as red arrow and the visually assigned trail vectors shown as white arrows. As in Figure~\ref{fig:mfmtk}, the peak of light is represented by a blue cross and the centre of light is represented by a red cross.

\begin{figure*}
    \centering
    \caption{\textbf{JClass 5}}
    \begin{minipage}{.23\textwidth}
        \centering
        \includegraphics[]{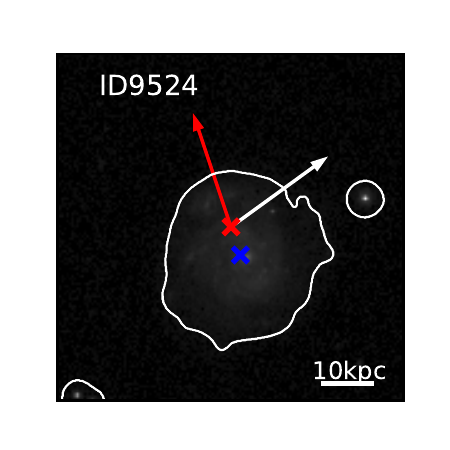}
    \end{minipage}
    \begin{minipage}{0.23\textwidth}
        \centering
        \includegraphics[]{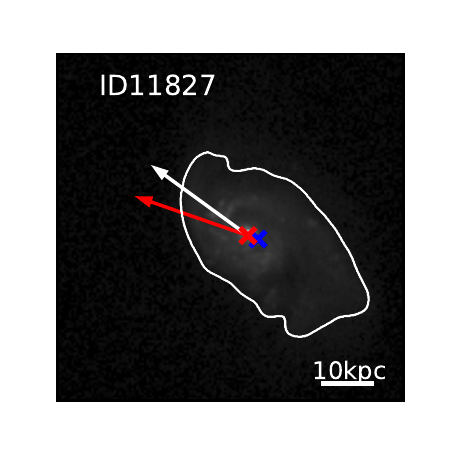}
    \end{minipage}
    \begin{minipage}{0.23\textwidth}
        \centering
        \includegraphics[]{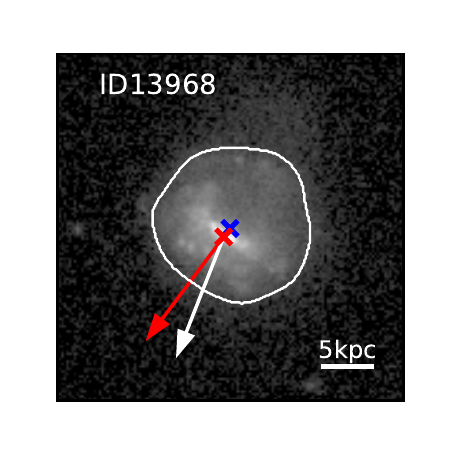}
    \end{minipage}
    \begin{minipage}{0.23\textwidth}
        \centering
        \includegraphics[]{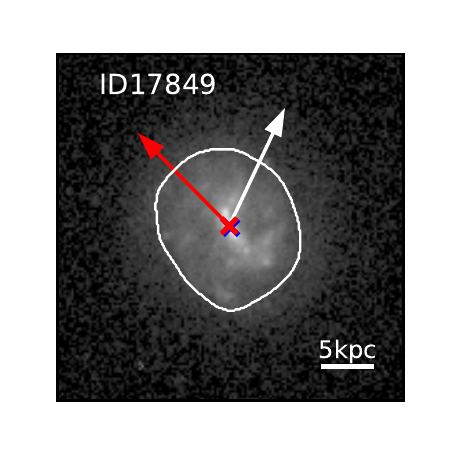}
    \end{minipage}
\end{figure*}

\begin{figure*}
\begin{minipage}{.23\textwidth}
    \centering
    \includegraphics[]{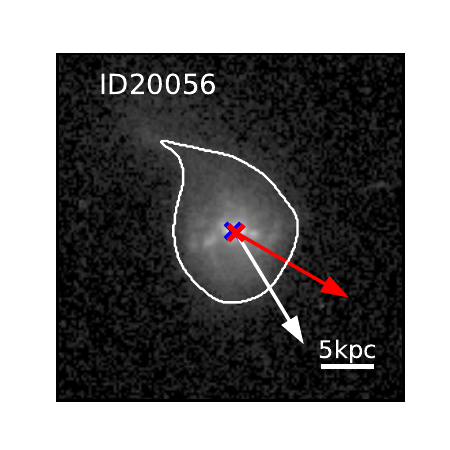}
\end{minipage}
\begin{minipage}{.23\textwidth}
    \centering
    \includegraphics{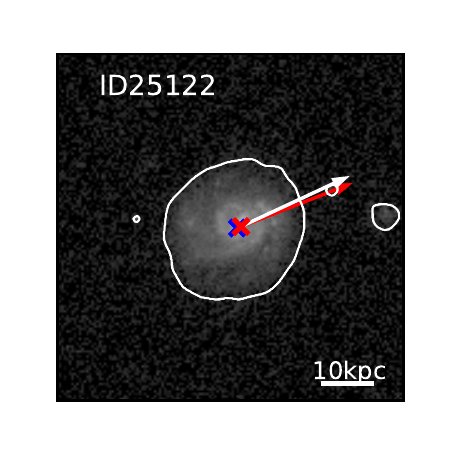}
\end{minipage}
\begin{minipage}{.23\textwidth}
    \centering
    \includegraphics{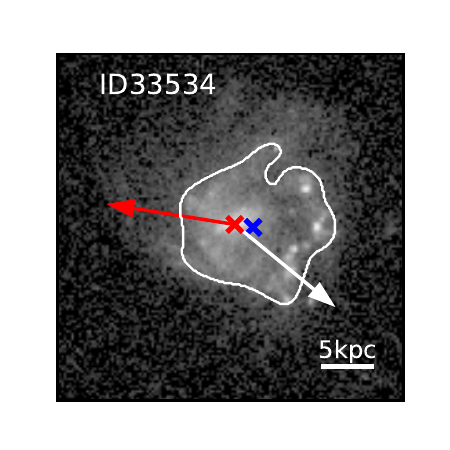}
\end{minipage}
\begin{minipage}{.23\textwidth}
    \centering
    \includegraphics{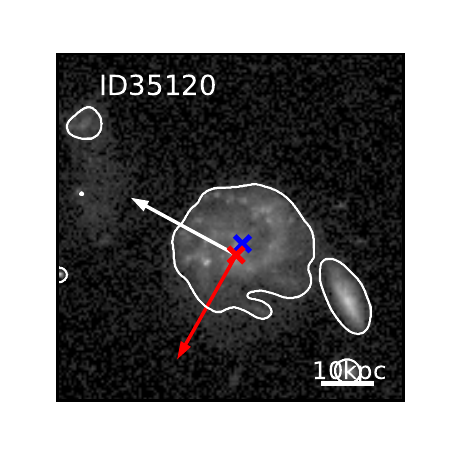}
\end{minipage}
\end{figure*}

\begin{figure*}
\begin{minipage}{.23\textwidth}
    \centering
    \includegraphics{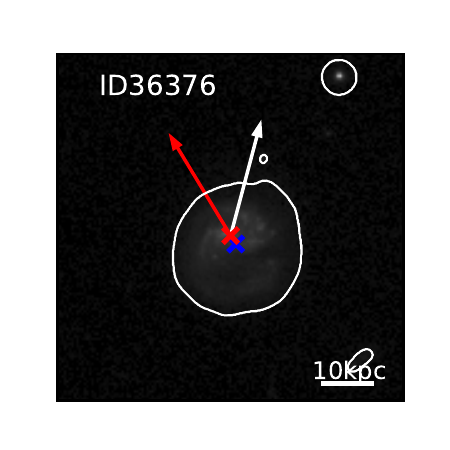}
\end{minipage}
\begin{minipage}{.23\textwidth}
    \centering
    \includegraphics{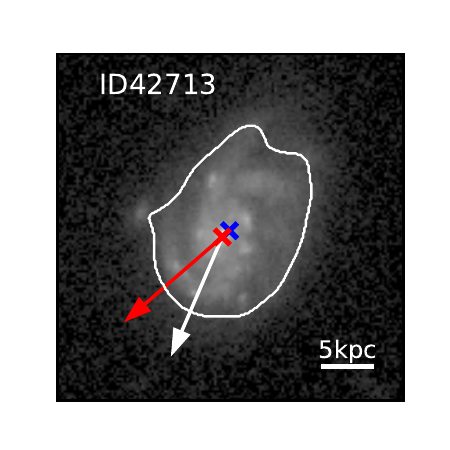}
\end{minipage}
\begin{minipage}{.23\textwidth}
    \centering
    \includegraphics{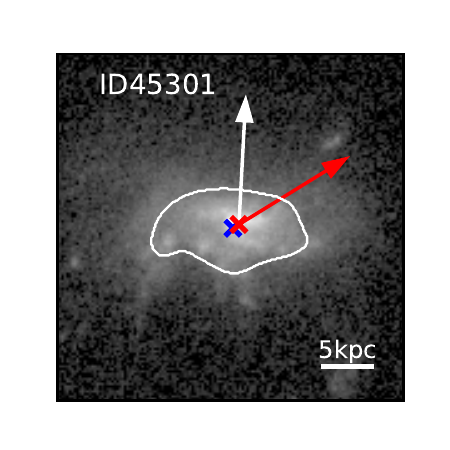}
\end{minipage}
\end{figure*}

%%%%%%%%%%%%%%%%%%%%%%%%%%%%%%%%%%%%
%%%%%%%%%%%%%%%%%%%%%%%%%%%%%%%%%%%%
%%%%%%%%%%%%%%%%%%%%%%%%%%%%%%%%%%%%

\begin{figure*}
\caption{\textbf{JClass 4}}
\begin{minipage}{.23\textwidth}
    \centering
    \includegraphics{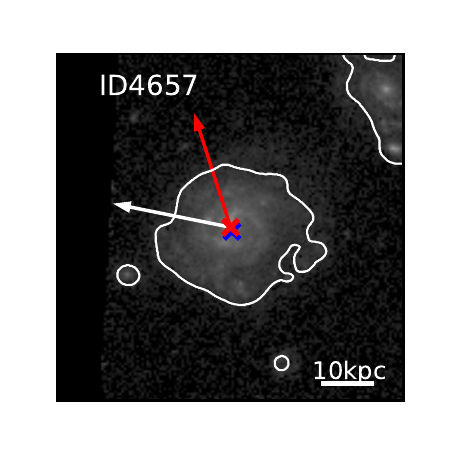}
\end{minipage}
\begin{minipage}{.23\textwidth}
    \centering
    \includegraphics{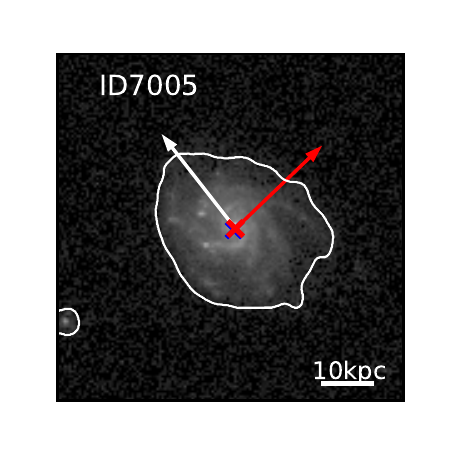}
\end{minipage}
\begin{minipage}{.23\textwidth}
    \centering
    \includegraphics{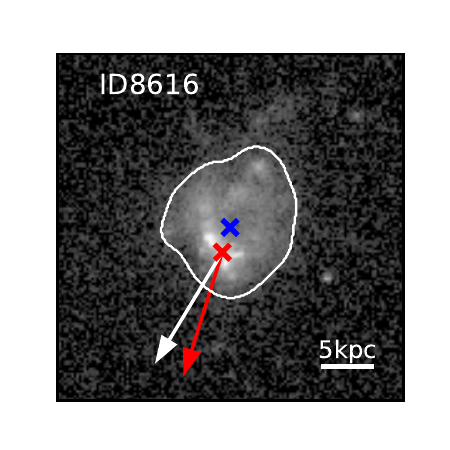}
\end{minipage}
\begin{minipage}{.23\textwidth}
    \centering
    \includegraphics{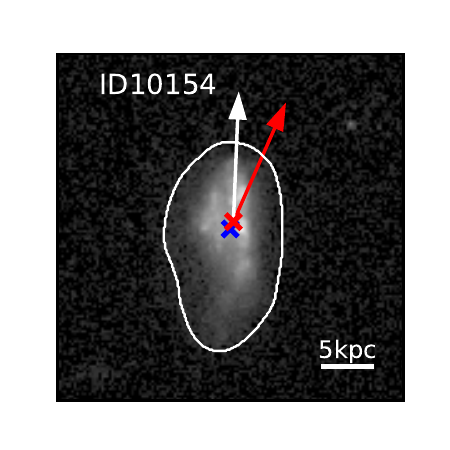}
\end{minipage}
\end{figure*}

\begin{figure*}
\begin{minipage}{.23\textwidth}
    \centering
    \includegraphics{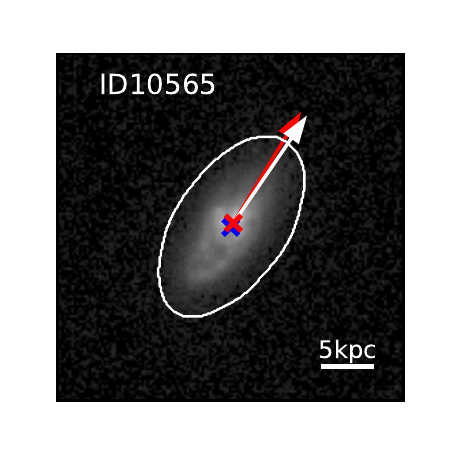}
\end{minipage}
\begin{minipage}{.23\textwidth}
    \centering
    \includegraphics{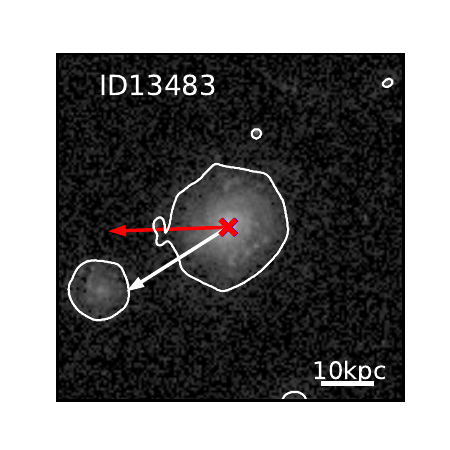}
\end{minipage}
\begin{minipage}{.23\textwidth}
    \centering
    \includegraphics{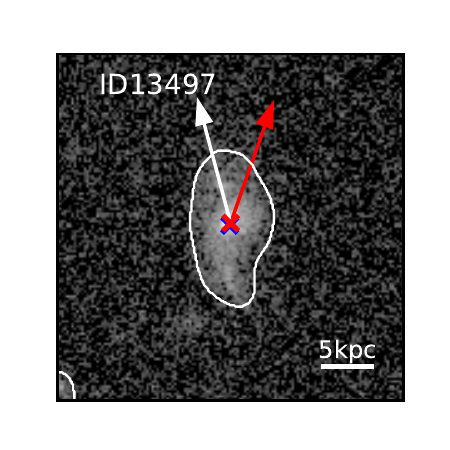}
\end{minipage}
\begin{minipage}{.23\textwidth}
    \centering
    \includegraphics{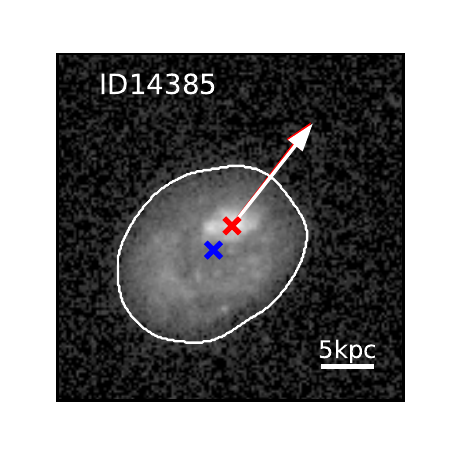}
\end{minipage}
\end{figure*}

\begin{figure*}
\begin{minipage}{.23\textwidth}
    \centering
    \includegraphics{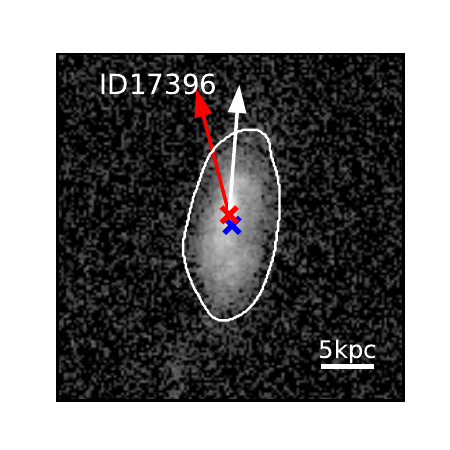}
\end{minipage}
\begin{minipage}{.23\textwidth}
    \centering
    \includegraphics{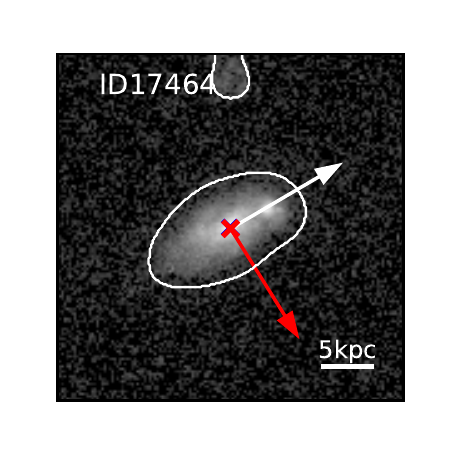}
\end{minipage}
\begin{minipage}{.23\textwidth}
    \centering
    \includegraphics{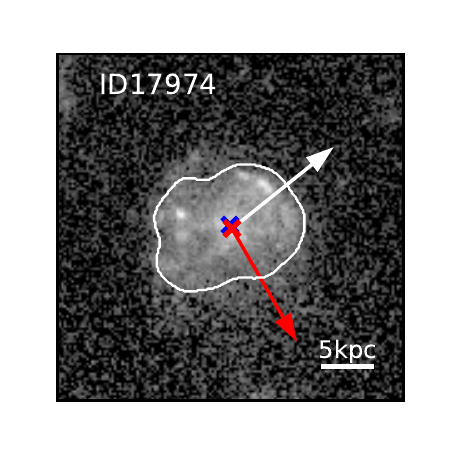}
\end{minipage}
\begin{minipage}{.23\textwidth}
    \centering
    \includegraphics{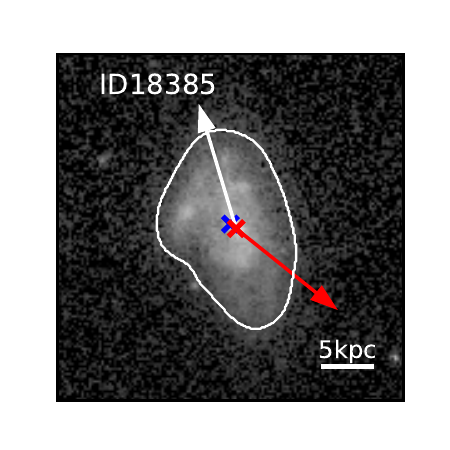}
\end{minipage}
\end{figure*}

\begin{figure*}
\begin{minipage}{.23\textwidth}
    \centering
    \includegraphics{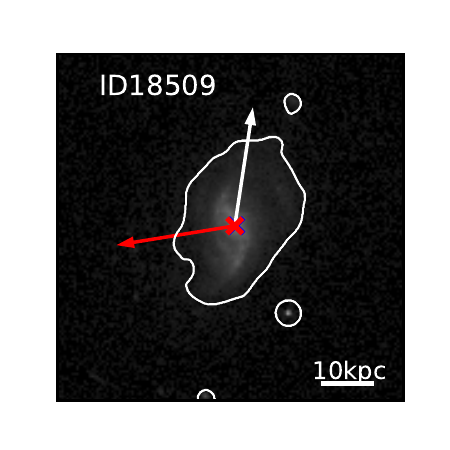}
\end{minipage}
\begin{minipage}{.23\textwidth}
    \centering
    \includegraphics{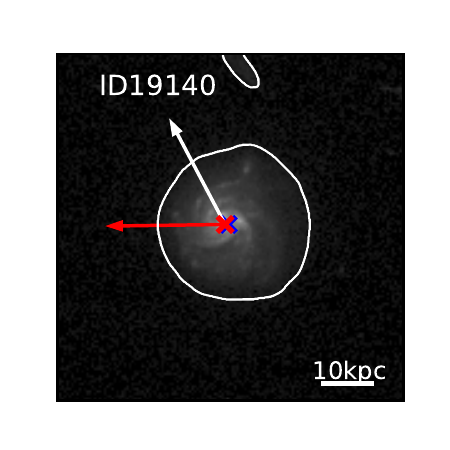}
\end{minipage}
\begin{minipage}{.23\textwidth}
    \centering
    \includegraphics{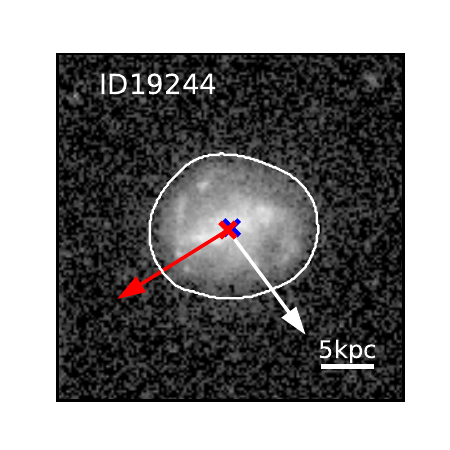}
\end{minipage}
\begin{minipage}{.23\textwidth}
    \centering
    \includegraphics{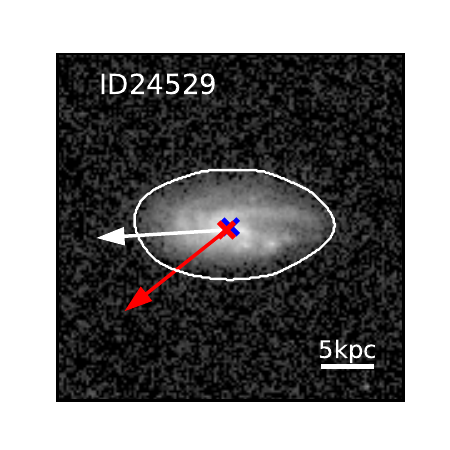}
\end{minipage}
\end{figure*}

\begin{figure*}
\begin{minipage}{.23\textwidth}
    \centering
    \includegraphics{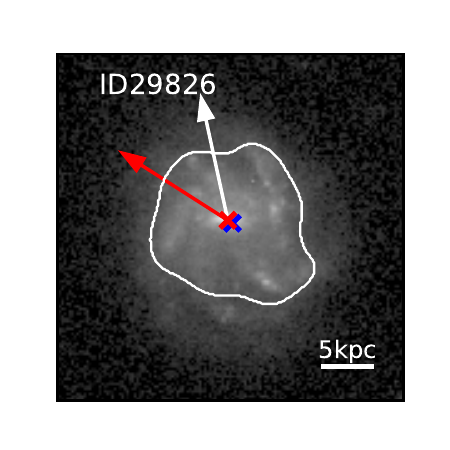}
\end{minipage}
\begin{minipage}{.23\textwidth}
    \centering
    \includegraphics{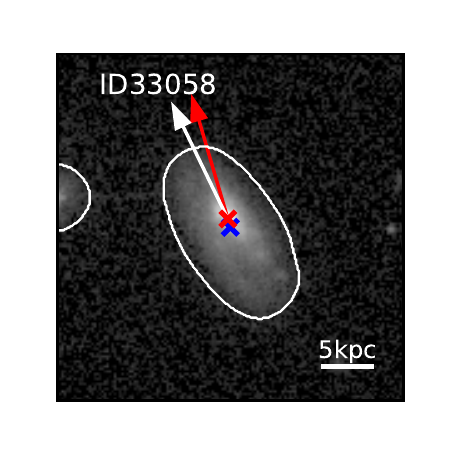}
\end{minipage}
\begin{minipage}{.23\textwidth}
    \centering
    \includegraphics{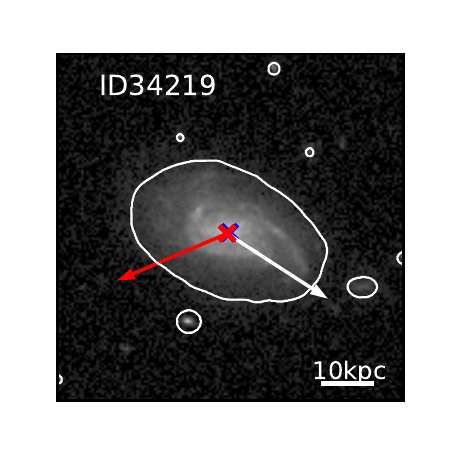}
\end{minipage}
\begin{minipage}{.23\textwidth}
    \centering
    \includegraphics{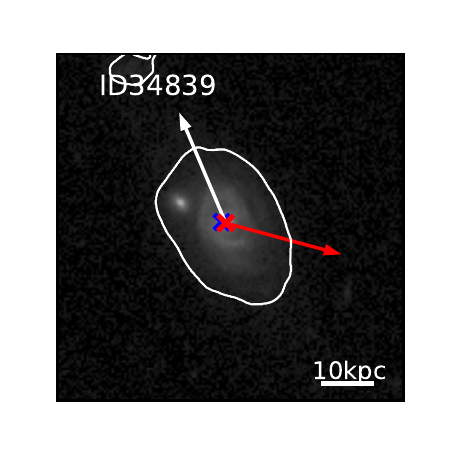}
\end{minipage}
\end{figure*}

\begin{figure*}
\begin{minipage}{.23\textwidth}
    \centering
    \includegraphics{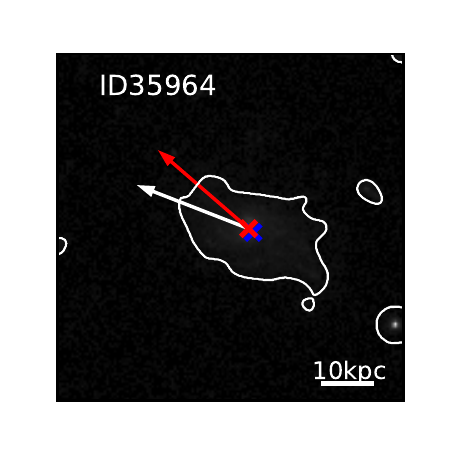}
\end{minipage}
\begin{minipage}{.23\textwidth}
    \centering
    \includegraphics{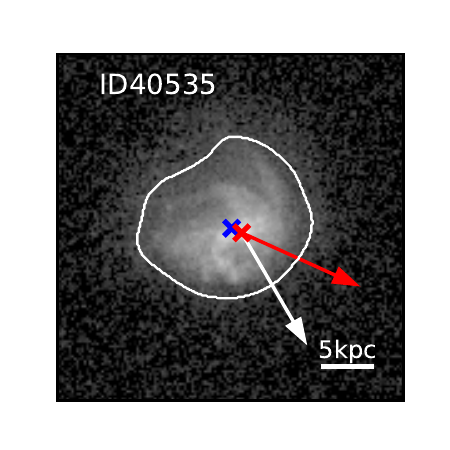}
\end{minipage}
\begin{minipage}{.23\textwidth}
    \centering
    \includegraphics{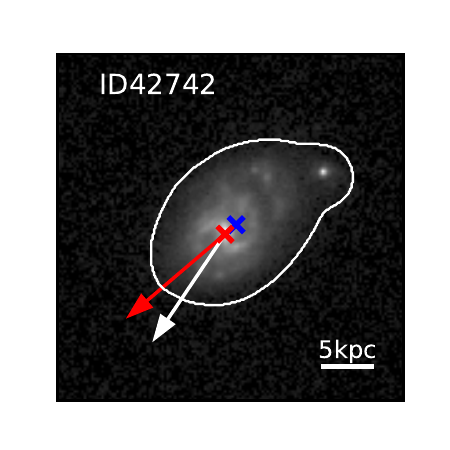}
\end{minipage}
\begin{minipage}{.23\textwidth}
    \centering
    \includegraphics{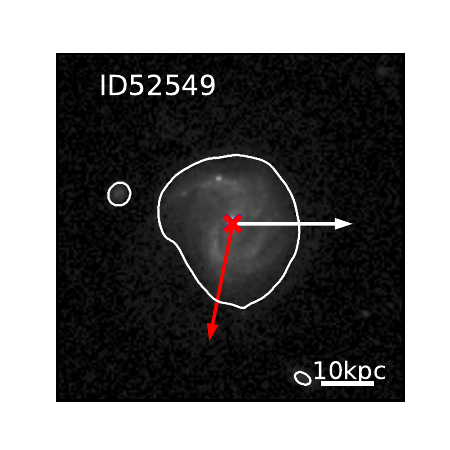}
\end{minipage}
\end{figure*}

%%%%%%%%%%%%%%%%%%%%%%%%%%%%%%%%%%%%
%%%%%%%%%%%%%%%%%%%%%%%%%%%%%%%%%%%%
%%%%%%%%%%%%%%%%%%%%%%%%%%%%%%%%%%%%

\begin{figure*}
\caption{\textbf{JClass 3}}
\begin{minipage}{.23\textwidth}
    \centering
    \includegraphics{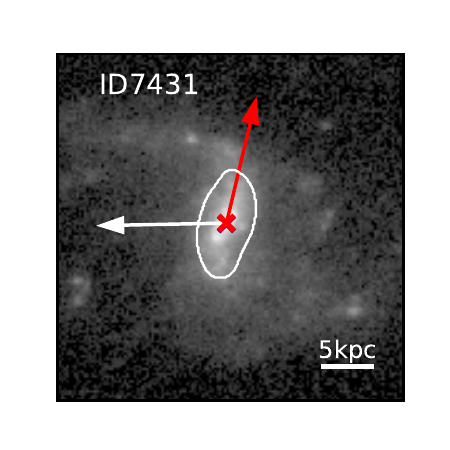}
\end{minipage}
\begin{minipage}{.23\textwidth}
    \centering
    \includegraphics{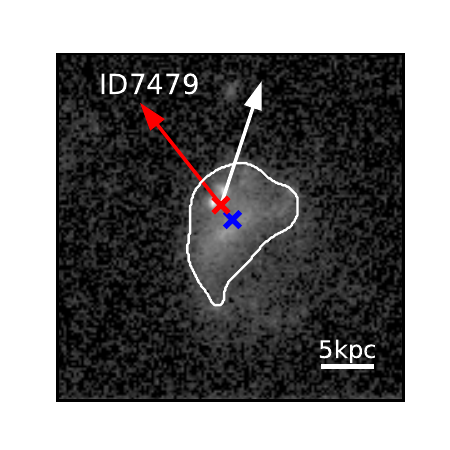}
\end{minipage}
\begin{minipage}{.23\textwidth}
    \centering
    \includegraphics{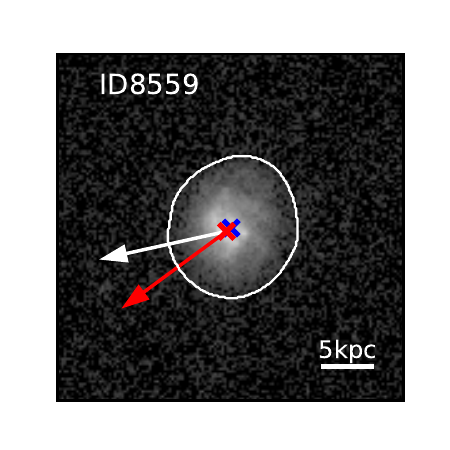}
\end{minipage}
\begin{minipage}{.23\textwidth}
    \centering
    \includegraphics{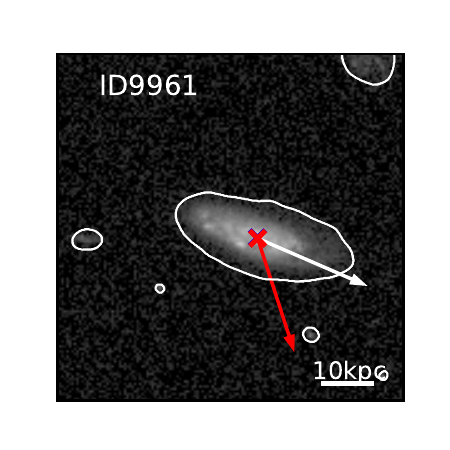}
\end{minipage}
\end{figure*}

\begin{figure*}
\begin{minipage}{.23\textwidth}
    \centering
    \includegraphics{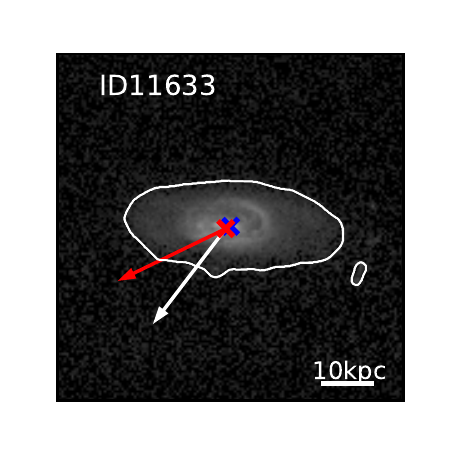}
\end{minipage}
\begin{minipage}{.23\textwidth}
    \centering
    \includegraphics{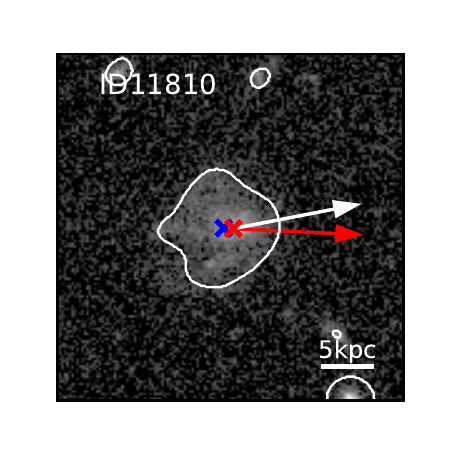}
\end{minipage}
\begin{minipage}{.23\textwidth}
    \centering
    \includegraphics{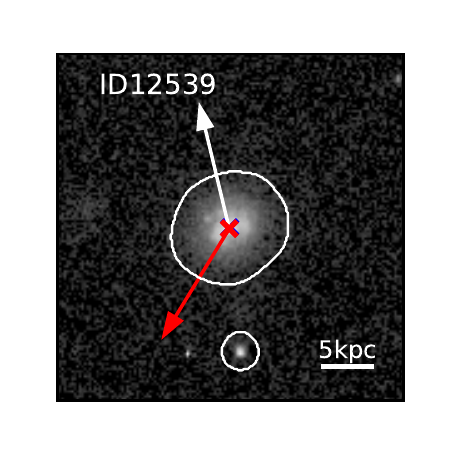}
\end{minipage}
\begin{minipage}{.23\textwidth}
    \centering
    \includegraphics{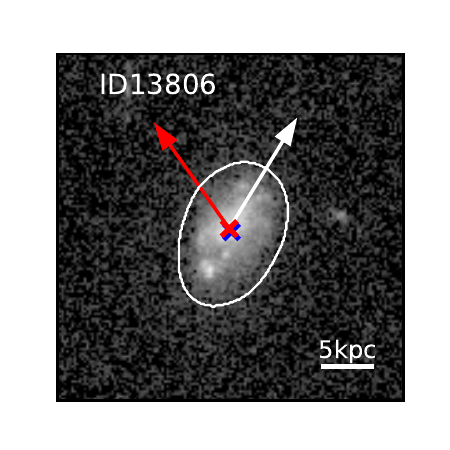}
\end{minipage}
\end{figure*}

\begin{figure*}
\begin{minipage}{.23\textwidth}
    \centering
    \includegraphics{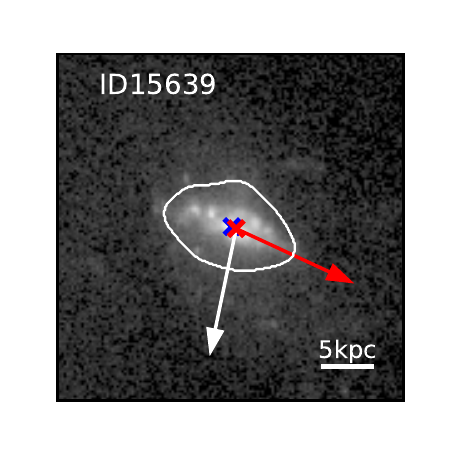}
\end{minipage}
\begin{minipage}{.23\textwidth}
    \centering
    \includegraphics{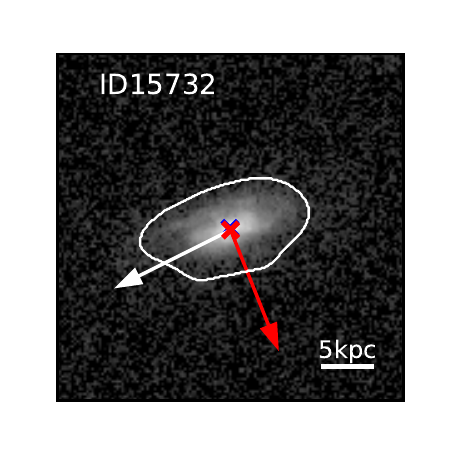}
\end{minipage}
\begin{minipage}{.23\textwidth}
    \centering
    \includegraphics{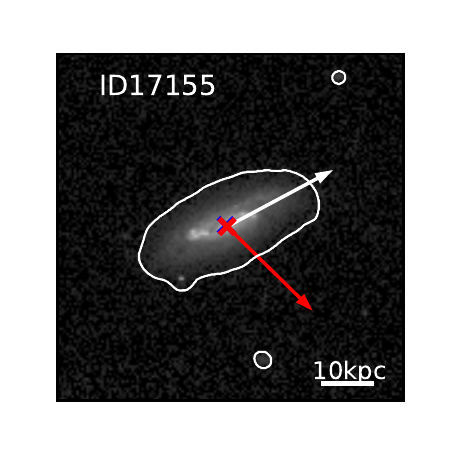}
\end{minipage}
\begin{minipage}{.23\textwidth}
    \centering
    \includegraphics{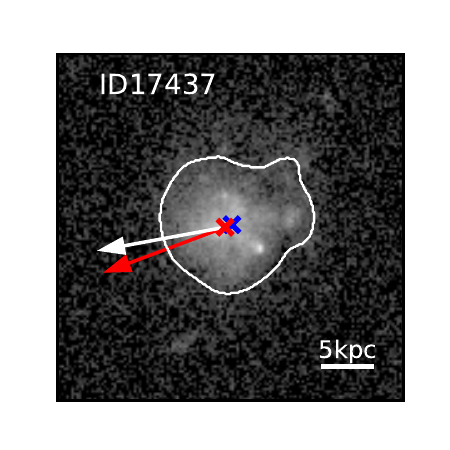}
\end{minipage}
\end{figure*}

\begin{figure*}
\begin{minipage}{.23\textwidth}
    \centering
    \includegraphics{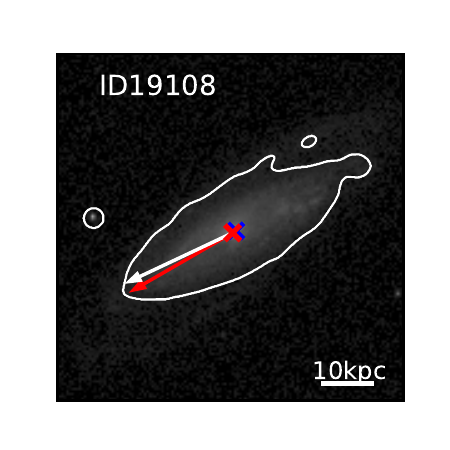}
\end{minipage}
\begin{minipage}{.23\textwidth}
    \centering
    \includegraphics{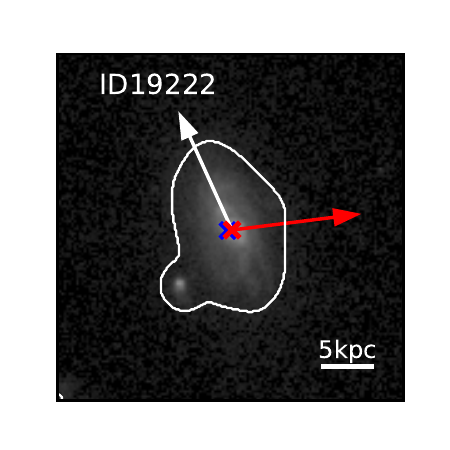}
\end{minipage}
\begin{minipage}{.23\textwidth}
    \centering
    \includegraphics{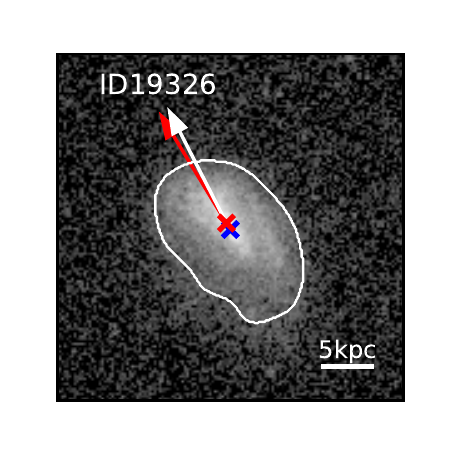}
\end{minipage}
\begin{minipage}{.23\textwidth}
    \centering
    \includegraphics{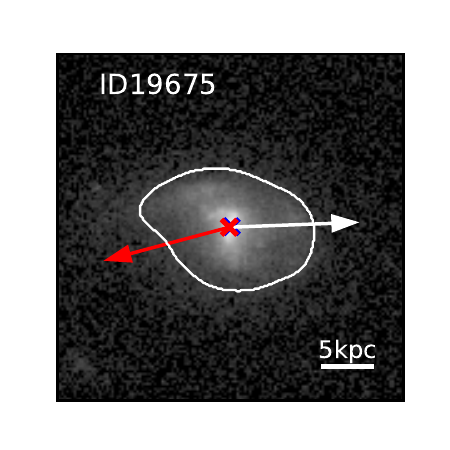}
\end{minipage}
\end{figure*}

\begin{figure*}
\begin{minipage}{.23\textwidth}
    \centering
    \includegraphics{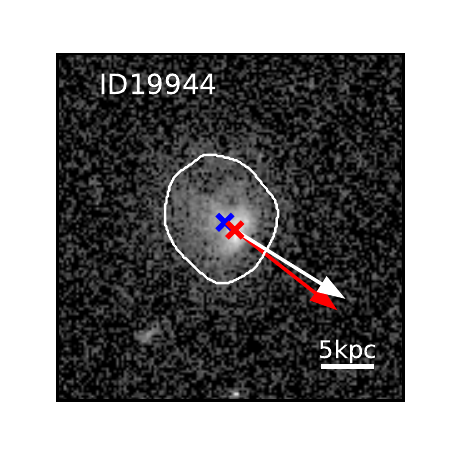}
\end{minipage}
\begin{minipage}{.23\textwidth}
    \centering
    \includegraphics{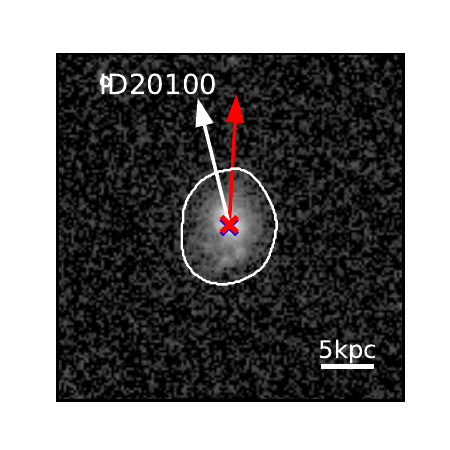}
\end{minipage}
\begin{minipage}{.23\textwidth}
    \centering
    \includegraphics{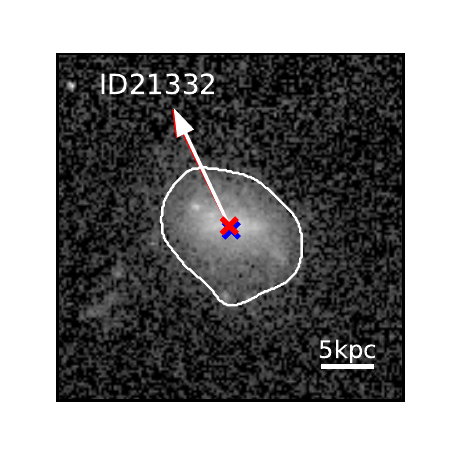}
\end{minipage}
\begin{minipage}{.23\textwidth}
    \centering
    \includegraphics{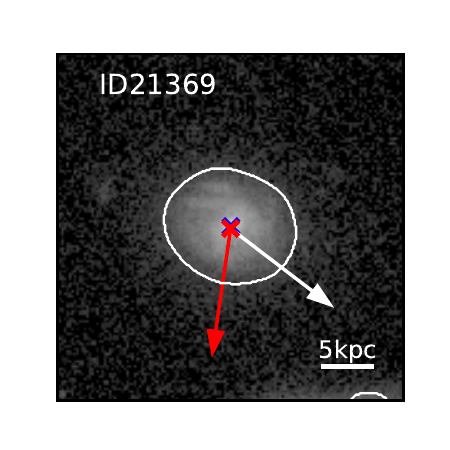}
\end{minipage}
\end{figure*}

\begin{figure*}
\begin{minipage}{.23\textwidth}
    \centering
    \includegraphics{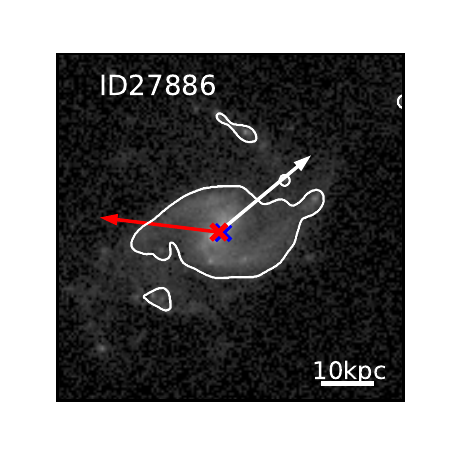}
\end{minipage}
\begin{minipage}{.23\textwidth}
    \centering
    \includegraphics{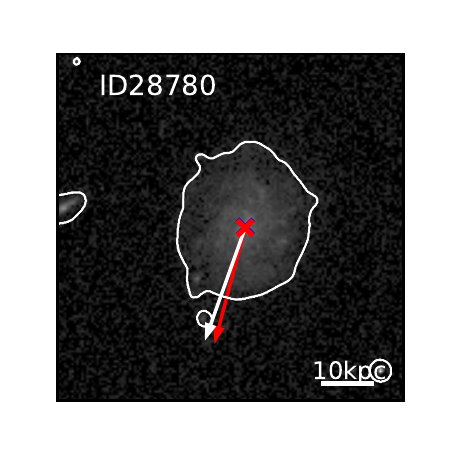}
\end{minipage}
\begin{minipage}{.23\textwidth}
    \centering
    \includegraphics{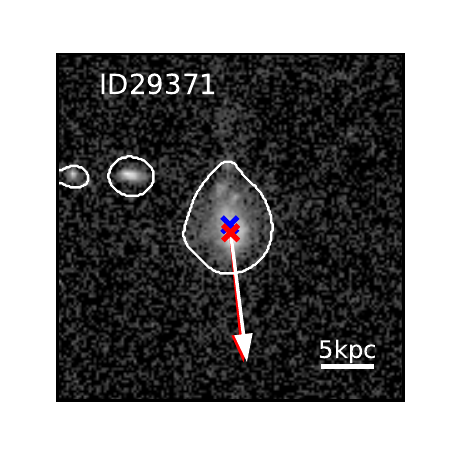}
\end{minipage}
\begin{minipage}{.23\textwidth}
    \centering
    \includegraphics{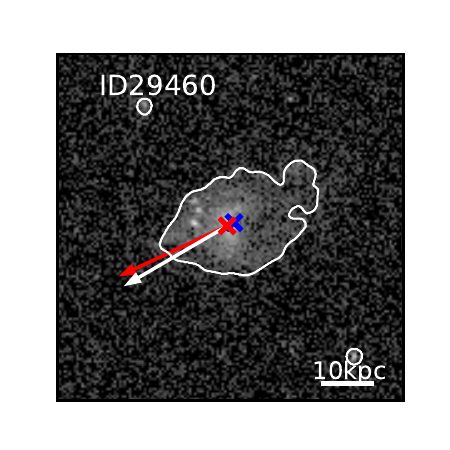}
\end{minipage}
\end{figure*}

\begin{figure*}
\begin{minipage}{.23\textwidth}
    \centering
    \includegraphics{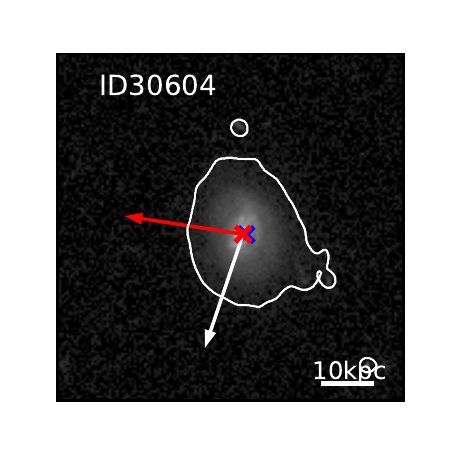}
\end{minipage}
\begin{minipage}{.23\textwidth}
    \centering
    \includegraphics{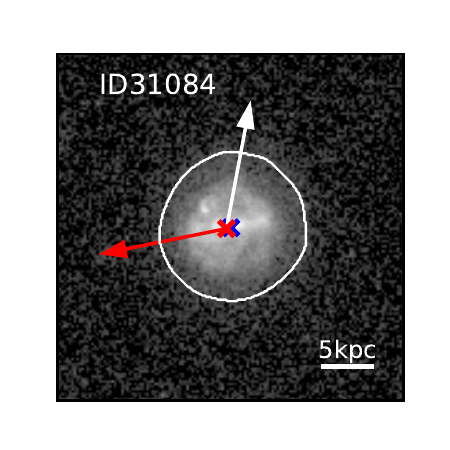}
\end{minipage}
\begin{minipage}{.23\textwidth}
    \centering
    \includegraphics{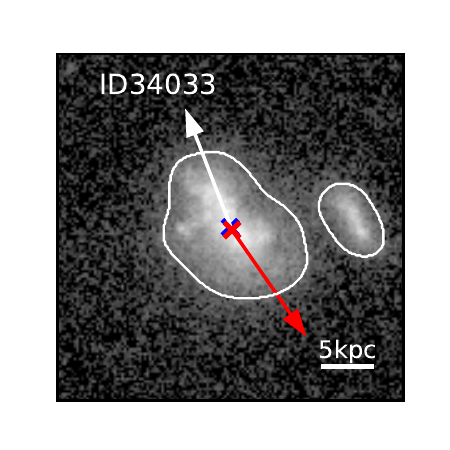}
\end{minipage}
\begin{minipage}{.23\textwidth}
    \centering
    \includegraphics{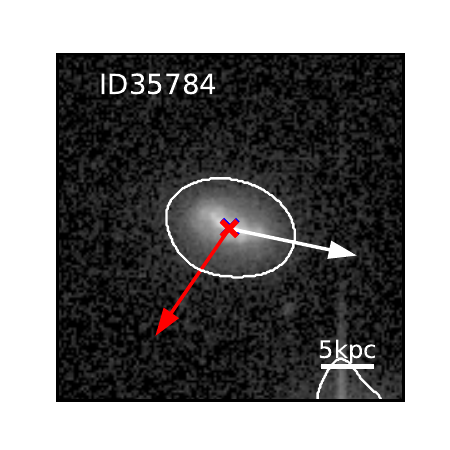}
\end{minipage}
\end{figure*}

\begin{figure*}
\begin{minipage}{.23\textwidth}
    \centering
    \includegraphics{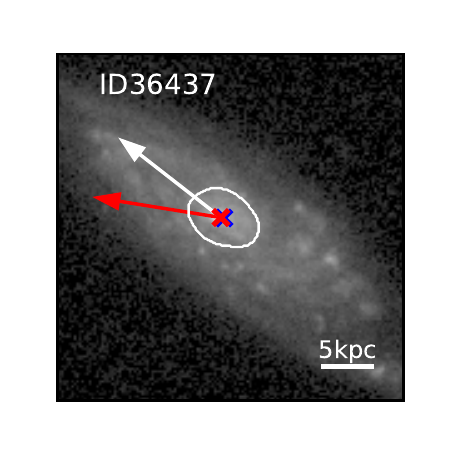}
\end{minipage}
\begin{minipage}{.23\textwidth}
    \centering
    \includegraphics{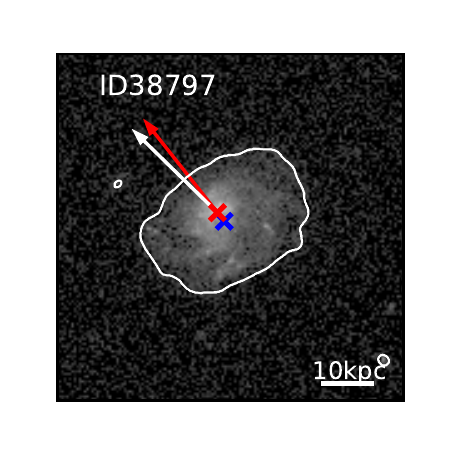}
\end{minipage}
\begin{minipage}{.23\textwidth}
    \centering
    \includegraphics{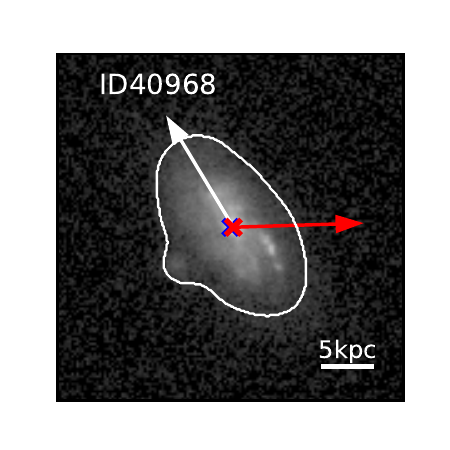}
\end{minipage}
\begin{minipage}{.23\textwidth}
    \centering
    \includegraphics{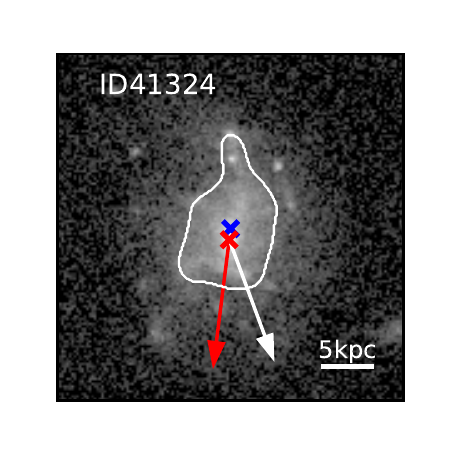}
\end{minipage}
\end{figure*}

\begin{figure*}
\begin{minipage}{.23\textwidth}
    \centering
    \includegraphics{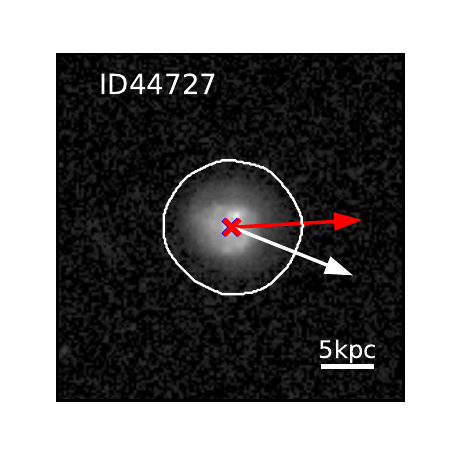}
\end{minipage}
\begin{minipage}{.23\textwidth}
    \centering
    \includegraphics{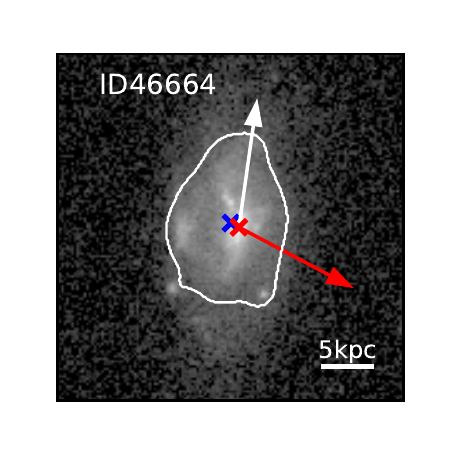}
\end{minipage}
\begin{minipage}{.23\textwidth}
    \centering
    \includegraphics{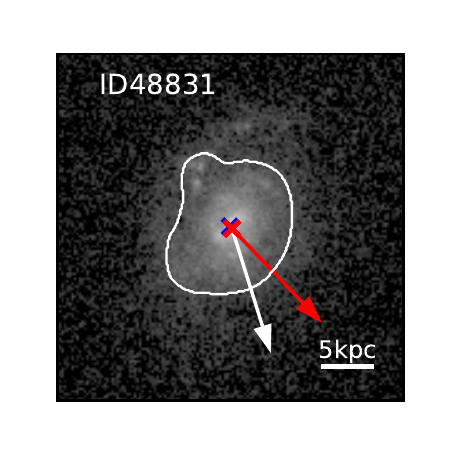}
\end{minipage}
\begin{minipage}{.23\textwidth}
    \centering
    \includegraphics{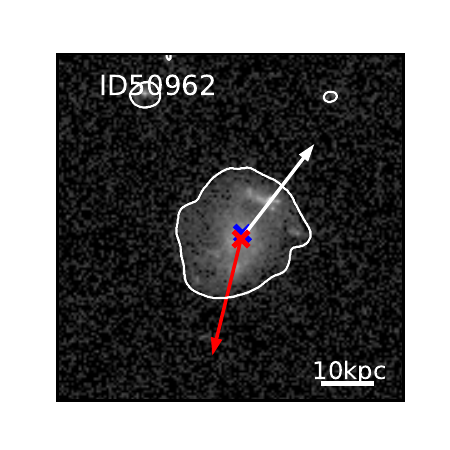}
\end{minipage}
\end{figure*}

\begin{figure*}
\begin{minipage}{.23\textwidth}
    \centering
    \includegraphics{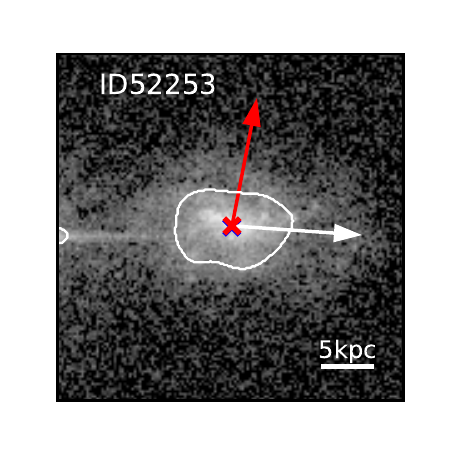}
\end{minipage}
\begin{minipage}{.23\textwidth}
    \centering
    \includegraphics{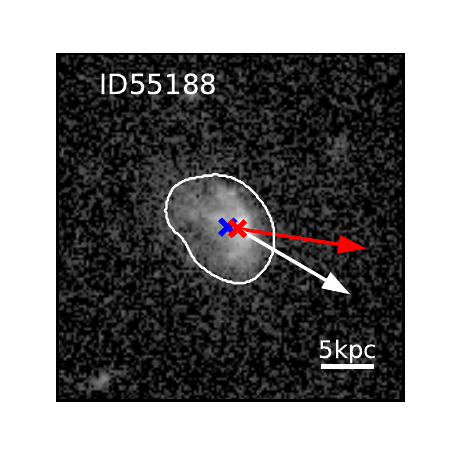}
\end{minipage}
\end{figure*}

%%%%%%%%%%%%%%%%%%%%%%%%%%%%%%%%%%%%%%%%%%%%%%%%%%

% Don't change these lines
\bsp	% typesetting comment
\label{lastpage}
\end{document}